\documentclass[showpacs,preprintnumbers,
floatfix]{revtex4}
\usepackage{graphicx}

\begin{document}

\preprint{YITP-03-22}

\title{Lattice Study of the Massive Schwinger Model with
$\theta$ Term under L\"{u}scher's ``Admissibility'' Condition}

\author{Hidenori Fukaya and Tetsuya Onogi}
\affiliation{Yukawa Institute for Theoretical Physics,
             Kyoto University, Kyoto 606-8502, Japan}



\begin{abstract}
L\"uscher's ``admissibility'' condition on the gauge field space 
plays an essential role in
constructing lattice gauge theories which has exact chiral 
symmetries. 
We apply the gauge action proposed by L\"uscher 
with the domain-wall fermion 
action to the numerical simulation of the massive Schwinger model.
We find this action can generate configurations in each
topological sector separately without any topology changes.
By developing a new method to sum over different topological sectors, 
we calculate meson masses in nonzero-$\theta$ vacuum.
\end{abstract}

\maketitle

\section{Introduction}

There are various problems in gauge theories where the 
chiral symmetry plays a crucial role. Although the lattice gauge
theories provide a method for nonperturbative 
computation on these problems, they have not given satisfactory 
answers since the conventional fermion action suffers 
from either the species doubling 
\cite{Nielsen:hk, Nielsen:1980rz, Nielsen:1981xu} or the
lack of chiral symmetry, which makes the study of the 
chiral behavior difficult.  For this reason, extensive studies 
have been made to improve the fermion action. It was found that
lattice Dirac operators satisfying the Ginsparg-Wilson relation
\cite{Ginsparg:1981bj}, 
which is for example realized by the Neuberger's overlap Dirac 
operator \cite{Neuberger:1997fp, Neuberger:1998wv},
have exact chiral symmetries\cite{Luscher:1998pq}.
Such actions are expected to give a fundamental improvement in the 
study of K meson physics, finite temperature QCD, or even chiral 
gauge theories, despite their complicated forms.

At the classical level, the Ginsparg-Wilson relation
is sufficient to solve the problem of chirality.
However, at the quantum level, topological properties 
of the gauge fields in the continuum space should also be
kept on the lattice in order to reproduce the 
correct chiral anomalies.
L\"uscher found that one can construct lattice gauge theories without
breaking topological structures by restricting the link variables
to satisfy the
``admissibility condition''\cite{Luscher:1998du} ;
\begin{equation}\label{eq:admi}
\|1-P_{\mu\nu}(x))\|<\epsilon \;\;\mbox{for all}\;\; x,\mu,\nu.
\end{equation}
where 
\begin{equation}
P_{\mu\nu}(x)=U_{\mu}(x)U_{\nu}(x+\hat{\mu}a)U^{\dagger}_{\mu}(x+\hat{\nu}a)
U^{\dagger}_{\nu}(x),
\end{equation}
and $\epsilon$ is a fixed positive number.
In order for the gauge fields to satisfy this condition
automatically, he proposed 
the following action as an example, 
\begin{eqnarray}
S_{G}=\left\{\begin{array}{ll}
\beta\displaystyle{\sum_{x, \mu > \nu}}\frac{(1-{\rm Re} P_{\mu\nu}(x))}
{1-\|1-P_{\mu\nu}(x)\|/\epsilon}
& \mbox{if} \;\;\|1-P_{\mu\nu}(x)\| <\epsilon
\\ \infty & \mbox{otherwise}
\end{array}\right. .
\label{eq:lataction}
\end{eqnarray}

The admissibility condition makes gauge fields smooth and unphysical
configurations such as vortices are suppressed.
In fact, the space of admissible fields is separated into 
disconnected subspaces labeled by some integers
which correspond to topological charges in the continuum
theory \cite{Luscher:1981zq, Luscher:1998kn}.
Therefore, one can precisely treat topological effects such as
U(1)problem, $\theta$ vacuum and so on. 
Moreover, L\"uscher's action is also differentiable and  gauge invariant 
and has a good continuum limit as Wilson's action.

Although there are other proposals for improvements 
of the gauge action \cite{Iwasaki,  Alford, DBW2, DBW2_2,CP-PACS,RBC,Jung}, 
they do not keep any topological structures.
In this sense L\"uscher's action combined with Ginsparg-Wilson fermion action
would be the best choice to investigate chiral symmetries 
on the lattice.
Moreover, admissibility condition is indispensable to
construct chiral gauge theories since the gauge symmetry is
never realized without exact chiral symmetries
\cite{Luscher:1998du}\cite{Luscher:1999un}.
Thus it would be important to examine how admissibility 
works and how topological structures are realized on the lattice 
in numerical simulations.

In this paper, we apply L\"uscher's gauge action 
to the numerical studies of the two-flavor massive 
Schwinger model \cite{Jaster:1996nk} on the lattice 
using the domain wall fermion action
\cite{Kaplan, Shamir} in order to examine how the admissibility 
works by probing the topological properties of the lattice theories. 
The massive Schwinger model is a good test ground for 
several reasons; it has been well examined analytically in 
continuum space even in strong coupling limit,
its vacuum has non trivial topological structures
due to the chiral anomaly,  and it also shares many other 
interesting properties with QCD such as U(1) problem and confinement.
There already exist extensive studies of topological structures of the 
massive Schwinger model have on the lattice with Wilson's gauge action 
in the literature \cite{ref:Gat97qc,ref:Gat97bm,ref:Gat99gt,ref:Joos,ref:Lang,ref:Sharpe,ref:Duerr1}, where our work provides an alternative 
lattice approach with L\"uscher's action. Although it will be
interesting subject to make a detailed comparison of the results 
with our method with previous calculations, we will leave it for the 
future publication and focus on the feasibility study of our method and 
the new results on $\theta$ vacuum in the present work.

We found that L\"uscher's gauge action can generate 
configurations in each topological sector separately.
We develop a new method to evaluate the observables 
in nonzero-$\theta$ vacuum by summing over those in different 
sectors with correct weights.
 (Our strategy is quite different from that of sampling other 
  sectors by enhancing topology changes 
  \cite{Elser:1996tb, deForcrand:1997fm, Elser:2001pe, Vranas:1997da}.)
We applied our method to the meson correlators and observe the 
$\theta$ dependence of the isotriplet meson mass.
We reproduce the well-known continuum results;  the isotriplet 
meson mass scaling as a function of the fermion mass  and the 
their $\theta$ dependence, and the fact that the isosinglet 
meson acquires a heavier mass than the isotriplet meson 
due to anomaly (the so-called U(1) problem).

This paper is organized as follows.
In section \ref{sec:Schwinger review}, 
we summarize main results of the continuum massive Schwinger model.
In section \ref{sec:calculation details}, we discuss details of our 
simulation. In section \ref{sec:Results}, we present the results and 
compare them with continuum theory. In Appendix, we compare
L\"uscher's action and Wilson's action in the quenched approximation,
and show the impact of the admissibility condition on the 
topological and chiral properties.

\section{Review of the massive Schwinger model}\label{sec:Schwinger review}

\subsection{Continuum theory}
 We consider the two flavor massive Schwinger model
\cite{Schwinger:tp,Coleman:1975pw,Coleman:1976uz} with degenerate masses.
The continuum action in Euclidean space is defined as 
\begin{eqnarray}
S &=& S_{G} + S_{F},\nonumber\\
S_{G}&=& \int d^{2}x \frac{1}{4g^{2}}F_{\mu\nu}(x)F^{\mu\nu}(x), \nonumber\\
S_{F}&=& \int d^{2}x \sum_{i=1}^{2}
\bar{\psi_{i}}(x)(D\mbox{\hspace{-.1in}}/+m)\psi_{i}(x),
\end{eqnarray}
where
\begin{eqnarray}
F_{\mu\nu}(x) = \partial_{\mu}A_{\nu}(x)-\partial_{\nu}A_{\mu}(x),
\;\;\;
D\mbox{\hspace{-.1in}}/
= \sum_{\mu=1}^{2}\gamma^{\mu}(\partial_{\mu}+iA_{\mu}),
\nonumber\\
\gamma^{1}=\left(\begin{array}{cc}
	  0&1\\
	  1&0 
       \end{array}\right),
\;\;
\gamma^{2}=\left(\begin{array}{cc}
	  0&-i\\
	  i&0
       \end{array}\right),
\;\;
\gamma^{3}=-i\gamma^{1}\gamma^{2},
\end{eqnarray}
$A_{\mu}$ is the gauge field and $\psi$ is the two-spinor fermion field.
We take $g$ and $m$ to be positive without losing generality.

If we take the space-time to be a torus $T^{2}$,
the space of gauge fields is separated into topological
sectors each of which is labeled by an integer
\begin{equation}\label{eq:topcharge}
N = \frac{1}{4\pi}\int_{T^{2}} d^{2}x \epsilon_{\mu\nu}F^{\mu\nu},
\end{equation}
where  we take the sign convention of the antisymmetric tensor as 
$\epsilon_{12}=1$.
Then this theory has vacuums dependent on  phase $\theta$.
Full path integrals are defined by a summation of integrals in each 
topological sector;
\begin{eqnarray}\label{eq:generating func}
Z_{full}(g,m,\theta)&=& \sum_{N=-\infty}^{+\infty}
e^{iN \theta}Z_{N}(g,m)\nonumber\\
Z_{N}(g,m)&=& \int DA^{N}_{\mu}D\bar{\psi}D\psi
\exp(-S_{G}-S_{F}),
\end{eqnarray}
where $A^{N}_{\mu}$ denote gauge fields in the sector with topological
charge $N$.
Using Eq.(\ref{eq:topcharge}), we can rewrite 
$Z_{full}(g,m,\theta)$ as follows
\begin{equation}
Z_{full}(g,m,\theta) = \int DA_{\mu}D\bar{\psi}D\psi
\exp(-S_{G}-S_{F}-S_{\theta}),
\end{equation}
where
\begin{equation}
S_{\theta} = - i\int d^{2}x \frac{\theta}{4\pi}\epsilon_{\mu\nu}F^{\mu\nu}.
\end{equation}
\\

It is well-known that this model is equivalent to 
the two-component  scalar theory 
\cite{Coleman:1975pw, Coleman:1976uz, Frohlich:mt}
\begin{eqnarray}
S&=&\int d^{2}x\left[
\frac{1}{2}\partial_{\mu}\phi_{+}(x)\partial^{\mu}\phi_{+}(x)
+\frac{1}{2}\partial_{\mu}\phi_{-}(x)\partial^{\mu}\phi_{-}(x)
+\frac{\mu_{0}^{2}}{2}(\phi_{+}(x))^{2} \right.
\nonumber\\
&&
\left. -2cmg
\cos \left(\sqrt{2\pi}\phi_{+}(x)-\frac{\theta}{2}\right)
\cos (\sqrt{2\pi}\phi_{-}(x))\right],
\end{eqnarray}
where $\mu_{0}=g \displaystyle{\sqrt{\frac{2}{\pi}}}$ and $c$ 
is a numerical constant.

For $m \ll \mu_{0}$ and $\theta \sim 0$,
perturbative calculations of $O(m)$  show that light 
scalar $\phi_{-}$ has the mass 
\begin{equation}\label{eq:light mass}
m_{-}=\sqrt{2\pi}(2cm\mu_{0}^{1/2}\cos \frac{\theta}{2})^{2/3},
\end{equation}
and heavy scalar $\phi_{+}$ has the mass 
\begin{equation}\label{eq:heavy mass}
m_{+}=\mu_{0} + (O(m)\; \mbox{corrections}).
\end{equation}

Now, we discuss the chiral behaviors of the two-flavor Schwinger model
in the chiral limit $m\to 0$.
The action has $U(2)_{A} \simeq SU(2)_{A}\times U(1)_{A}$ chiral
symmetry in this limit. 
$U(1)_A$ symmetry is broken by the anomaly, which manifests
itself in the vacuum with nontrivial topological structures.
We define the isotriplet meson operator
$\pi_{0} \equiv \bar{\psi_{1}}\gamma_{3}\psi_{1}
-\bar{\psi_{2}}\gamma_{3}\psi_{2}$ 
and the isosinglet meson operator 
$\eta \equiv \bar{\psi_{1}}\gamma_{3}\psi_{1}+
\bar{\psi_{2}}\gamma_{3}\psi_{2}$
by the fermion bilinears. In the bosonization picture,
it is shown that $\pi_{0}$ propagation corresponds to 
that of light scalar $\phi_{-}$  and 
$\eta$ propagation corresponds to that of heavy scalar $\phi_{+}$.
In the massless limit, Eqs. (\ref{eq:light mass}) and 
(\ref{eq:heavy mass}) show that  $\pi_{0}$ becomes massless while 
$\eta$ remains massive in accordance with the 
U(1) problem in two-dimensional QED.

\subsection{Lattice theory}\label{sec:lattice reguralization}

Let us consider the lattice regularization of the 
massive Schwinger model.
We take the lattice size is $L\times L \times L_{3}$ 
with lattice spacing $a=1$, where $L_{3}$ is the length
of the third direction for the domain wall fermions \cite{Kaplan, Shamir}.
Action is defined as follows
\begin{eqnarray}
S &=& \beta S_{G} + S_{F},\\
S_{G}&=& \left\{\begin{array}{ll}
\displaystyle{\sum_{P}}\frac{(1-{\rm Re}P_{\mu\nu}(x))}
{1-(1-{\rm Re}P_{\mu\nu}(x))/\epsilon}
& \mbox{if admissible}
\\ \infty & \mbox{otherwise}
\end{array}
\right. ,\label{eq:admaction2}\\
S_{F} &=& \sum_{x,x^{\prime}}\sum_{s,s^{\prime}}
\sum^{2}_{i=1}
\left[
\bar{\psi}^{i}_{s}(x)D_{DW}(x,s ; x^{\prime},s^{\prime})
\psi^{i}_{s^{\prime}}(x^{\prime})\right.
\nonumber\\
&&
\left. + \phi^{i\ast}_{s}(x)D_{AP}(x,s ; x^{\prime},s^{\prime} )
\phi^{i}_{s^{\prime}}(x^{\prime})
\right],
\end{eqnarray}
where
\begin{eqnarray}
D_{DW}(x,s ; x^{\prime},s^{\prime})
&=&
\frac{1}{2}\sum^{2}_{\mu=1}\left\{
(1+\gamma_{\mu})U_{\mu}(x)\delta_{x+\hat{\mu},x^{\prime}}
\delta_{s,s^{\prime}}\right.
\nonumber\\
&&
\left. +(1-\gamma_{\mu})U^{\dagger}_{\mu}(x-\hat{\mu})
\delta_{x-\hat{\mu},x^{\prime}}\delta_{s,s^{\prime}}
\right\}
\nonumber\\
&&
+(M-3)\delta_{x,x^{\prime}}\delta_{s,s^{\prime}}
\nonumber\\
&&
+P_{+}\delta_{s+1,s^{\prime}}\delta_{x,x^{\prime}}
+P_{-}\delta_{s-1,s^{\prime}}\delta_{x,x^{\prime}}
\nonumber\\
&&
+(m-1)P_{+}\delta_{s,L_{3}}\delta_{s^{\prime},1}\delta_{x,x^{\prime}}
+(m-1)P_{-}\delta_{s,1}\delta_{s^{\prime},L_{3}}\delta_{x,x^{\prime}}
,\nonumber\\
D_{AP}(x,s ; x^{\prime},s^{\prime})
&=&
\frac{1}{2}\sum^{2}_{\mu=1}\left\{
(1+\gamma_{\mu})U_{\mu}(x)\delta_{x+\hat{\mu},x^{\prime}}
\delta_{s,s^{\prime}}\right.
\nonumber\\
&&
\left. +(1-\gamma_{\mu})U^{\dagger}_{\mu}(x-\hat{\mu})
\delta_{x-\hat{\mu},x^{\prime}}\delta_{s,s^{\prime}}
\right\}
\nonumber\\
&&
+(M-3)\delta_{x,x^{\prime}}\delta_{s,s^{\prime}}
\nonumber\\
&&
+P_{+}\delta_{s+1,s^{\prime}}\delta_{x,x^{\prime}}
+P_{-}\delta_{s-1,s^{\prime}}\delta_{x,x^{\prime}}
\nonumber\\
&&
-2P_{+}\delta_{s,L_{3}}\delta_{s^{\prime},1}\delta_{x,x^{\prime}}
-2P_{-}\delta_{s,1}\delta_{s^{\prime},L_{3}}\delta_{x,x^{\prime}}.
\end{eqnarray}
 $\beta = 1/g^{2}$, $M$ is a constant satisfying $0<M<1$,
$\displaystyle{\sum_{P}}$ denotes summation over all plaquettes and
$P_{\pm}$ are the chiral projection operators;
\begin{equation}
P_{\pm}=\frac{1 \pm \gamma_{3}}{2}.
\end{equation}
$m$ is the fermion mass.
$\phi^{i}$'s are Pauli-Villars regulators which cancel the bulk 
contribution.

Since it is not possible to change the topological charge by 
a local updation under the admissibility condition, 
our lattice theory with L\"uscher's gauge  action
has a topological invariant 
\begin{equation}\label{eq:topcharge2}
N = -\frac{i}{2\pi}\sum_{x} \ln P_{12}(x).
\end{equation}
This charge corresponds to Eq.(\ref{eq:topcharge}) and
gauge field configurations are classified into topological sectors .
Each sector characterized by $N$ has the classical gauge 
configuration $U^{cl}_{\mu}(x,y)$ minimizing the action, which is 
given as  
\begin{eqnarray}\label{eq:const background}
U^{cl[N]}_{1}(x,y) &=& \exp\left\{\frac{2 \pi i \nu_1}{L} -\frac{2\pi Ni}{L}
\delta_{x,L-1} x \right\},
\nonumber\\
U^{cl[N]}_{2}(x,y) &=& \exp\left\{\frac{2 \pi i \nu_2}{L} + 
\frac{2\pi Ni}{L^{2}} y\right\},
\end{eqnarray}
up to gauge transformations, where $\nu_1$ and $\nu_2$ are 
the parameters which determine the values of Wilson lines 
in $x$ and $y$ directions.
$\nu_{1,2}$ can take any values in the region $0 \leq \nu_{1,2} < 1$ .
This configuration gives constant background 
electric fields over the torus.

\section{Lattice simulations}\label{sec:calculation details}

\subsection{Observables in each sector}\label{sec:parameters}

Simulation is carried out by Hybrid Monte Carlo method with
L\"uscher's gauge action Eq.(\ref{eq:admaction2}).
The matrix inversions are calculated by the conjugate gradient algorithm.

\begin{figure}[t]
\includegraphics[height=8cm,angle=-90]{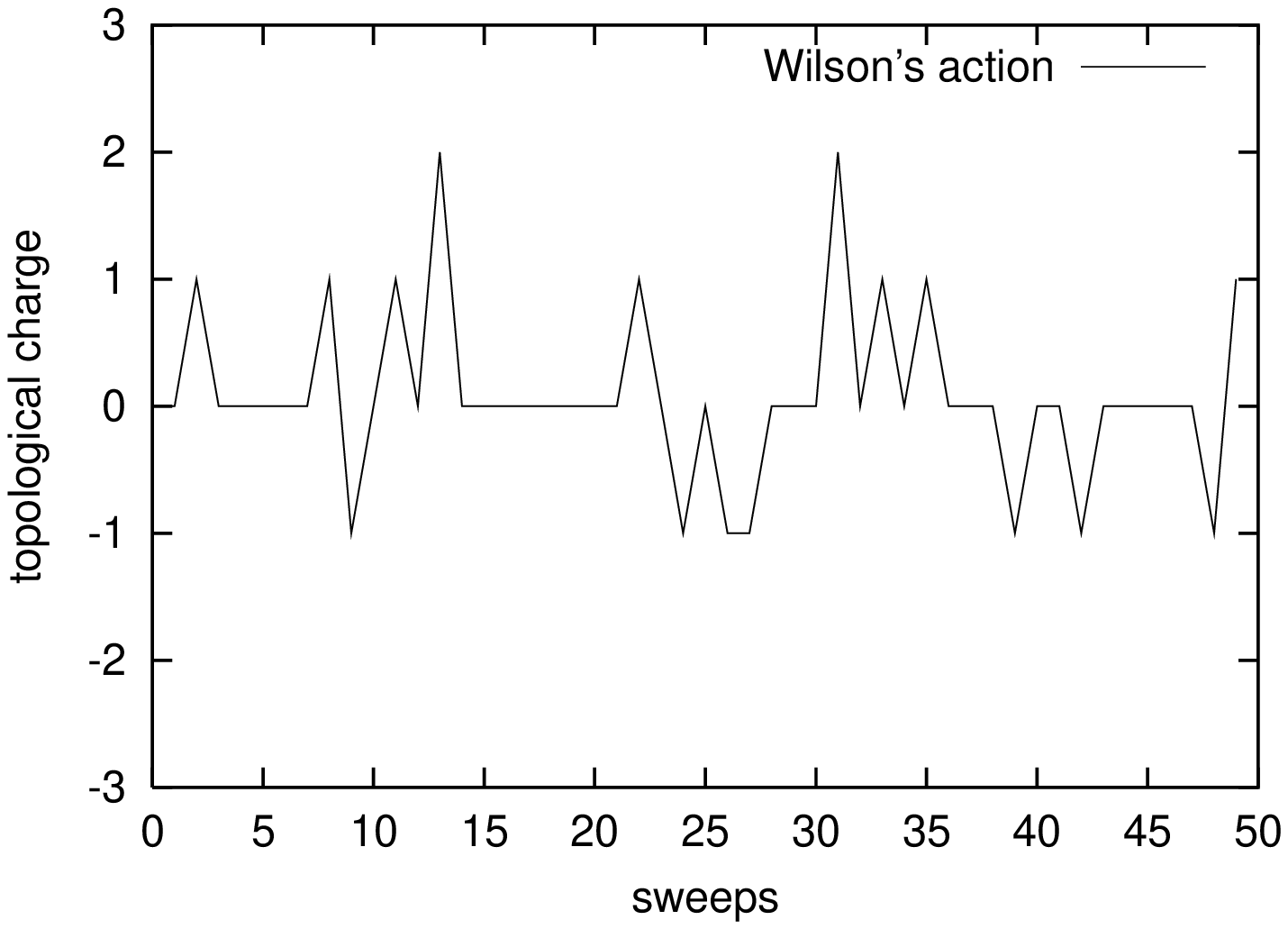}
\includegraphics[height=8cm,angle=-90]{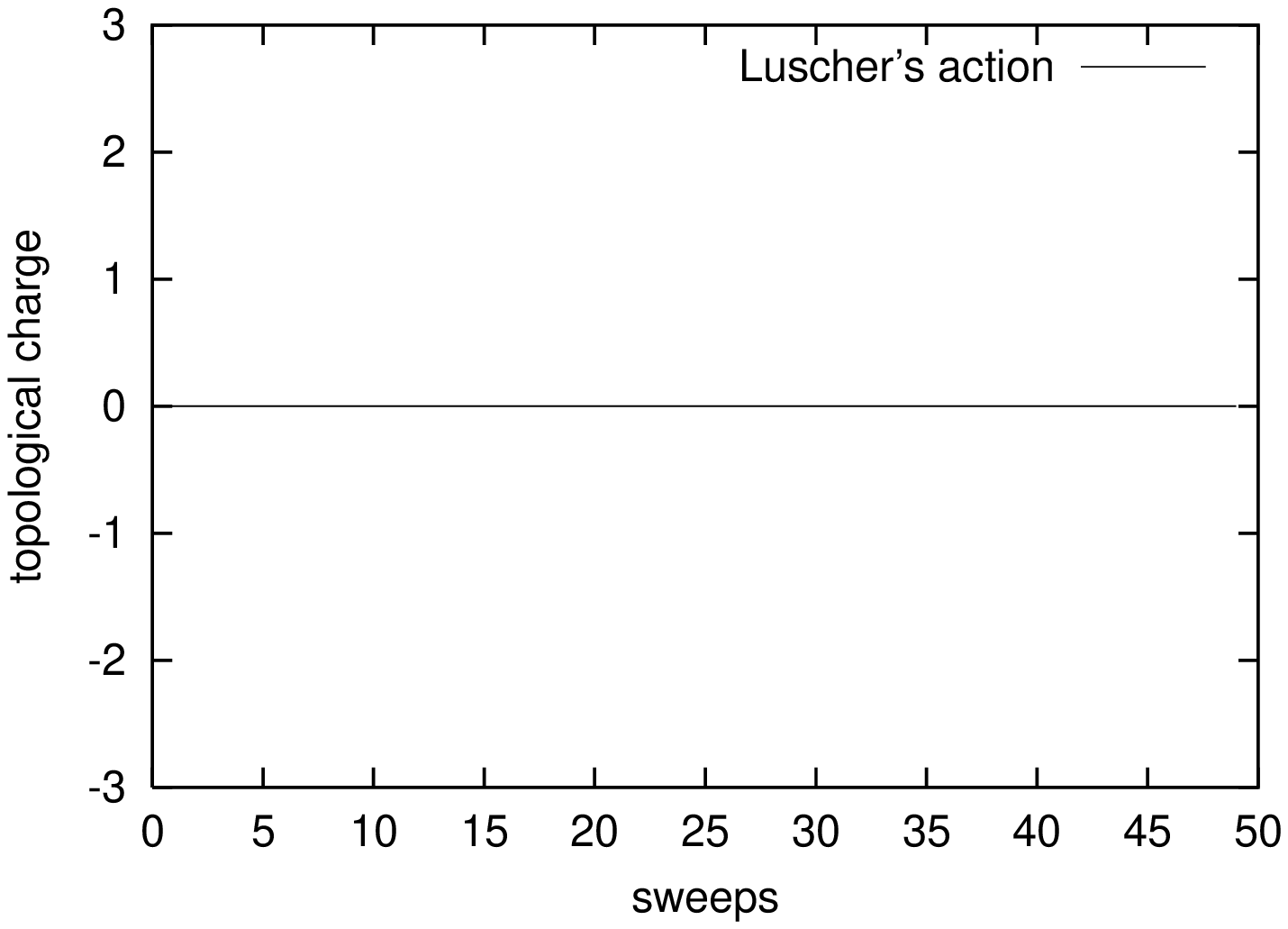}
\caption{The comparison of the Monte Carlo evolution of the
topological charge with Wilson's gauge action and L\"uscher's gauge
action for the same lattice spacings determined from the string tension.
Initial topological charges are zero. 
Left: Wilson's gauge action with $\beta=2.0$, Right: L\"uscher's gauge 
action with $\beta=0.5$. The topological charge changes
for Wilson's gauge action, while it does not for L\"uscher's action.}
\label{fig:top1}
\includegraphics[height=8cm,angle=-90]{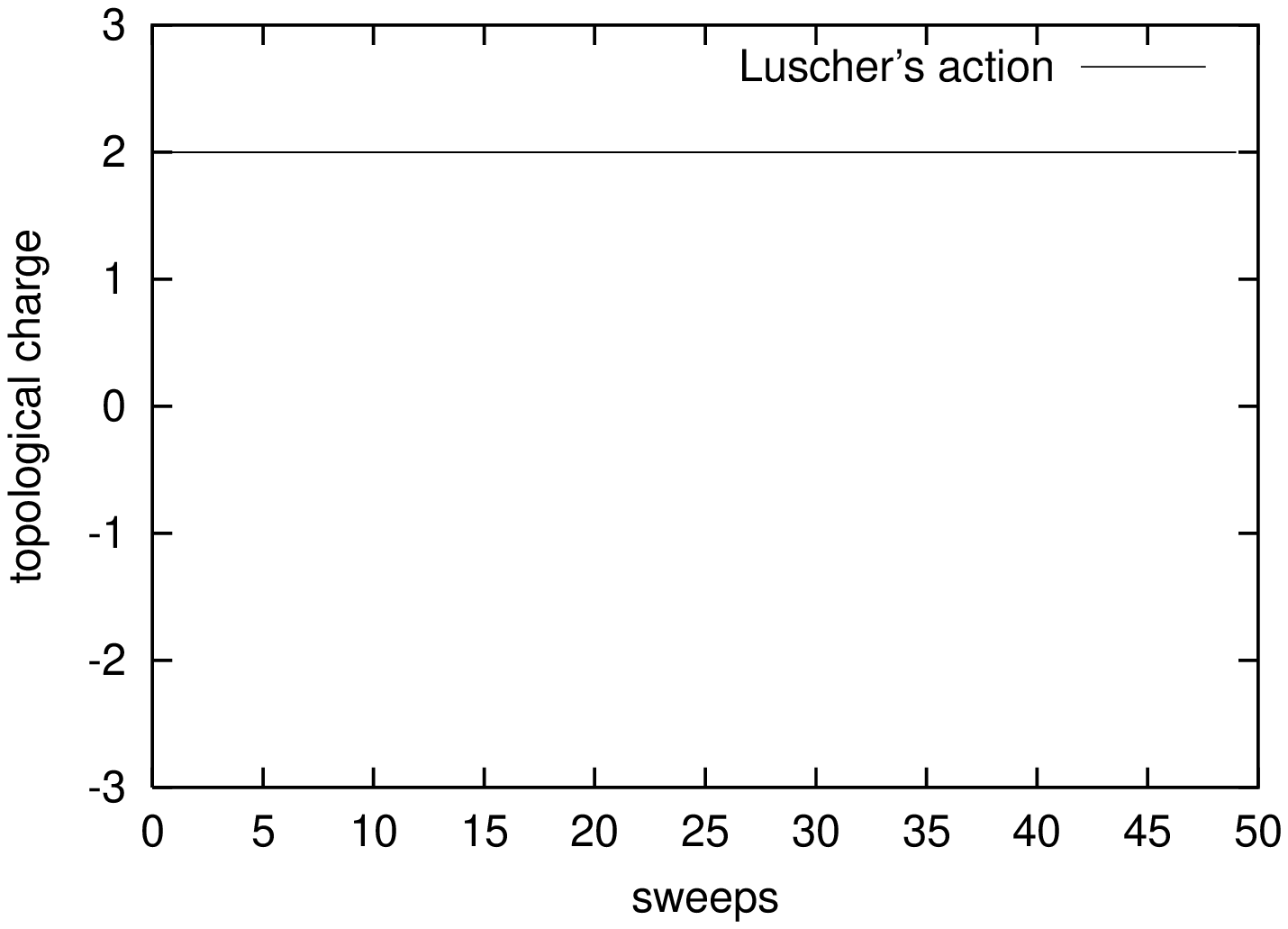}
\caption{The Monte Carlo evolution of the topological charge with 
L\"uscher's action for $\beta=0.5$ is shown. The initial topological 
charge is two. }
\label{fig:top2}
\end{figure}

We take $16\times 16\times 6$ lattice. 
The simulation is carried out at
$\beta = 1/g^{2} = 0.5$ and $M=0.9$. 
The parameter for the admissibility condition is chosen as  $\epsilon =
1.0$. At this value of $\epsilon$, we find that 
 initial topological charge is not changed through the simulation. 
( See Fig.~\ref{fig:top1} and Fig.~\ref{fig:top2}. )
For the fermion mass, we choose $m=0.1,0.2,0.3,0.4$.
50 molecular dynamics steps with stepsize $\Delta \tau =0.02$ 
are performed in one trajectory of the Hybrid Monte Carlo algorithm.
Configurations are updated per 10 trajectories.  
We generate 500 configurations for each topological sector by 
taking the classical configuration in Eq.(\ref{eq:const background}) 
as the initial configuration. 
From the set of configurations in each sector with topological charge 
$N$, we measure the isotriplet meson propagator 
\begin{equation}
C_{\pi}(x) = \displaystyle{\sum_{y}}\langle \pi(x,y)\pi(0,0)\rangle^{N}_{\beta,m},
\end{equation}
and the isosinglet meson propagator 
\begin{equation}
C_{\eta}(x) = \displaystyle{\sum_{y}}\langle \eta(x,y)\eta(0,0)\rangle^{N}_{\beta,m},
\end{equation}
where $\langle \rangle^{N}_{\beta,m}$ denotes the expectation value in the $N$ sector.

\newpage
\subsection{A new method of summing over different topological sectors}

The Hybrid Monte Carlo simulation is performed by small changes of 
link variables. Thus choosing the configuration 
given by Eq.(\ref{eq:const background}) as initial condition,
we can generate configurations without changing the topological charge
for any value of the coupling constant.

Now, we discuss full path integrals on $\theta$-vacuum.
Suppose that we measure the expectation value of an operator $O$,
\begin{equation}
\langle  O \rangle^{full}_{\beta,m} =  
\frac{\sum_{N=-\infty}^{+\infty}e^{iN \theta}
       \int DU^{N}_{\mu}D\bar{\psi}D\psi \;O e^{-\beta S_{G}-S_{F}}}
     {\sum_{N=-\infty}^{+\infty}e^{iN \theta} Z_N(\beta,m) }
\end{equation}
where $U^{N}_{\mu}$ denote link variables in the sector with $N$
and 
\begin{equation}
Z_N(\beta,m)=\int DU^{N}_{\mu}D\bar{\psi}D\psi e^{-\beta S_{G}-S_{F}} 
\end{equation}
is the lattice counter part of the $Z_N$ in Eq.(\ref{eq:generating func}).
In terms of the expectation values in each topological sector,
$\langle  O \rangle^{full}_{\beta,m}$ can be rewritten as,
\begin{equation}
\langle  O \rangle^{full}_{\beta,m} = 
\frac{ \sum_{N=-\infty}^{+\infty}e^{iN \theta}
       \langle  O \rangle^{N}_{\beta,m}R^{N}(\beta,m) }
     { \sum_{N=-\infty}^{+\infty}e^{iN \theta}
       R^{N}(\beta,m) },
\end{equation}
where
\begin{equation}
\langle  O \rangle^{N}_{\beta,m} = 
\frac{\int DU^{N}_{\mu}D\bar{\psi}D\psi\; O e^{-\beta S_{G}-S_{F}}}
     {Z_{N}(\beta,m)},
\end{equation}
and
\begin{equation}
R^{N}(\beta,m)=\frac{Z_{N}(\beta,m)}{Z_{0}(\beta,m)}.
\end{equation}
We call $R^{N}(\beta,m)$ the reweighting factor.
Note that $Z_{N}(\beta,m)$ satisfies the following 
differential equation;
\begin{equation}
\frac{\partial Z_{N}(\beta,m)}{\partial \beta}/Z_{N}(\beta,m)
= -\langle  S_{G} \rangle^{N}_{\beta,m}.
\end{equation}
By integrating over $\beta$ again, $Z_{N}(\beta,m)$ is 
expressed as,
\begin{equation}
Z_{N}(\beta,m)=Z_{N}(\infty,m)
\exp \left(\int_{\beta}^{\infty}d\beta^{\prime}
\langle  S_{G} \rangle^{N}_{\beta^{\prime},m}\right).
\end{equation}
Then, the reweighting factor $R^{N}(\beta,m)$ is 
expressed as,
\begin{eqnarray}
R^{N}(\beta,m)
&=&\frac{Z_{N}(\infty,m)}{Z_{0}(\infty,m)}
\exp \left[\int_{\beta}^{\infty}d\beta^{\prime}
\left(\langle  S_{G} \rangle^{N}_{\beta^{\prime},m}
-\langle  S_{G} \rangle^{0}_{\beta^{\prime},m}\right)\right]
\nonumber\\
&=&
\exp(-\beta S^{N}_{G\;min})
\frac{\int d\nu_1 d\nu_2 \det(D^{N}_{DW})^{2}/\det(D^{N}_{AP})^{2}}
    {\int d\nu_1 d\nu_2 \det(D^{0}_{DW})^{2}/\det(D^{0}_{AP})^{2}}
\nonumber\\
&&
\times \exp \left[\int_{\beta}^{\infty}d\beta^{\prime}
\left(\langle  S_{G} - S^{N}_{G\;min}\rangle^{N}_{\beta^{\prime},m}
-\langle  S_{G} \rangle^{0}_{\beta^{\prime},m}\right)\right],
\nonumber\\
\label{eq:reweight}
\end{eqnarray}
where $S^{N}_{G\;min}$ is minimum of the gauge action in sector with $N$ 
given by constant background fields 
Eq.(\ref{eq:const background}) and $D^{N}_{DW}$ and $D^{N}_{AP}$
are Dirac operators given by this background.
Note that the integrand vanishes rapidly as 
$\beta^{\prime}\to \infty$, so the integral 
over $\beta^{\prime}$ converges.

We can evaluate the full path integrals on $\theta$-vacuum
by obtaining $\langle  O \rangle^{N}_{\beta,m}$ and $R^{N}(\beta,m)$ in 
each sector.
It should be noted that this method is only possible 
with a gauge action with the admissibility condition 
in which the topological charge is strictly conserved.
In our approach we take the property that L\"uscher's gauge 
action allows no topology change at all as an advantage
and treat the sum over the topologies in a controlled fashion.
This is in contrast to the conventional gauge actions, 
with which the topology change is suppressed but not completely 
prohibited so that one has to tackle the problem of enhancing 
topology changes.

A related but somewhat different approach was proposed by 
D\"urr \cite{ref:Duerr2} where one makes a quenched calculation and give the whole 
fermion determinant as the reweighting factor. He also proposed an 
approximation in which one replaces the determinant for the given 
configuration by the determinant of a common representative configuration 
for the given sector, which reduces the enormous computational effort .

Of course, computing the reweighting factors for the sum over 
topologies requires extra works.  
Whether this program works must be examined in practical simulations.
In the following subsections we show that our new method is valid 
and full path integrals can be evaluated with 
controlled statistical and systematic errors.

\subsection{Calculation of $R^{N}(\beta,m)$}\label{sec:reweight}

\begin{figure}[b]
\includegraphics[height=8cm,angle=-90]{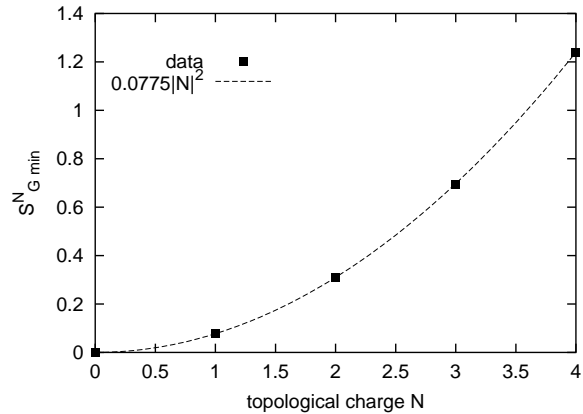}
\caption{ Minimum action in each topological sector is plotted 
as a function of the topological charge.
Filled squares are the data and the dotted line is 
the fit with a quadratic function.
}
\label{fig:free action}
\end{figure}

Let us discuss how to evaluate $R^{N}(\beta,m)$.
The classical minima of the gauge action 
$S^{N}_{G\;min}$ are evaluated easily. In Fig.~\ref{fig:free action},
we can see $S^{N}_{G\;min}$ is numerically proportional to $|N|^{2}$.

\begin{figure}[t]
\includegraphics[height=8cm,angle=-90]{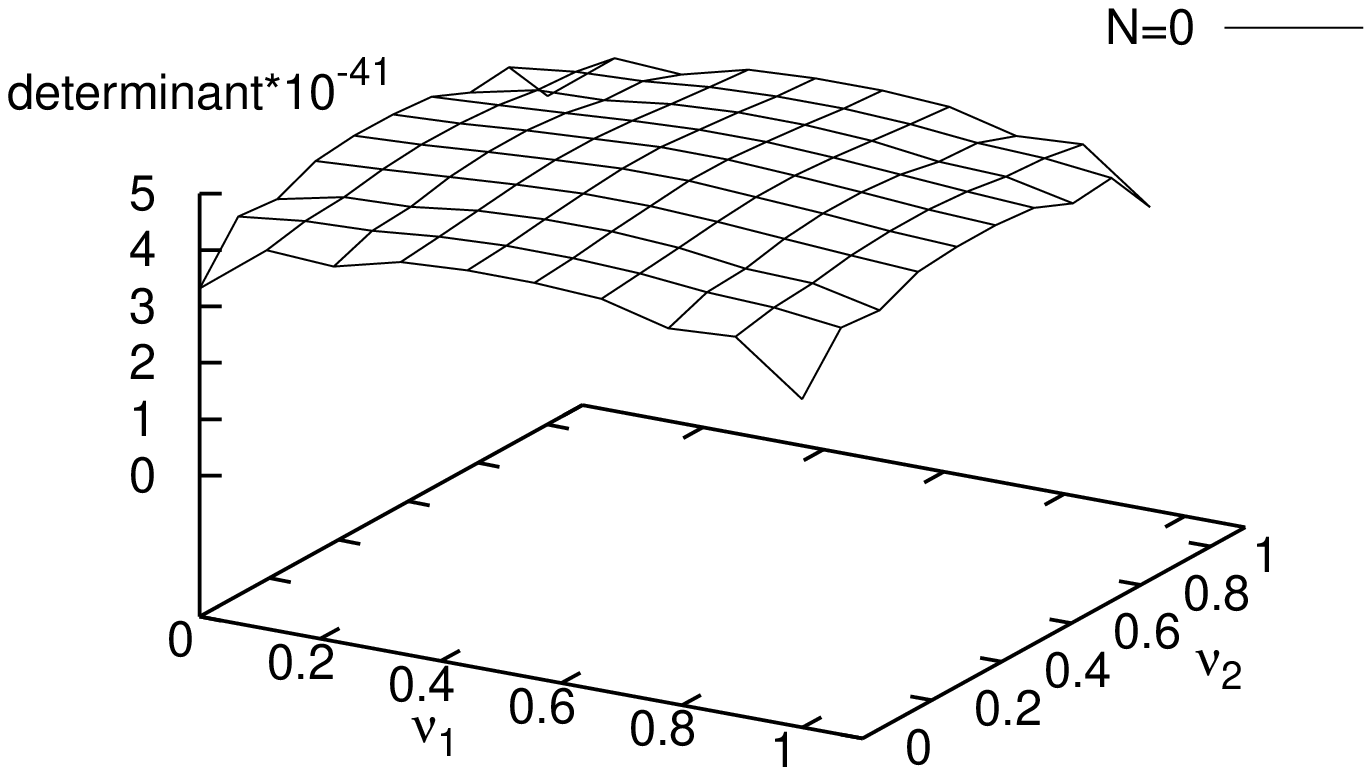}
\includegraphics[height=8cm,angle=-90]{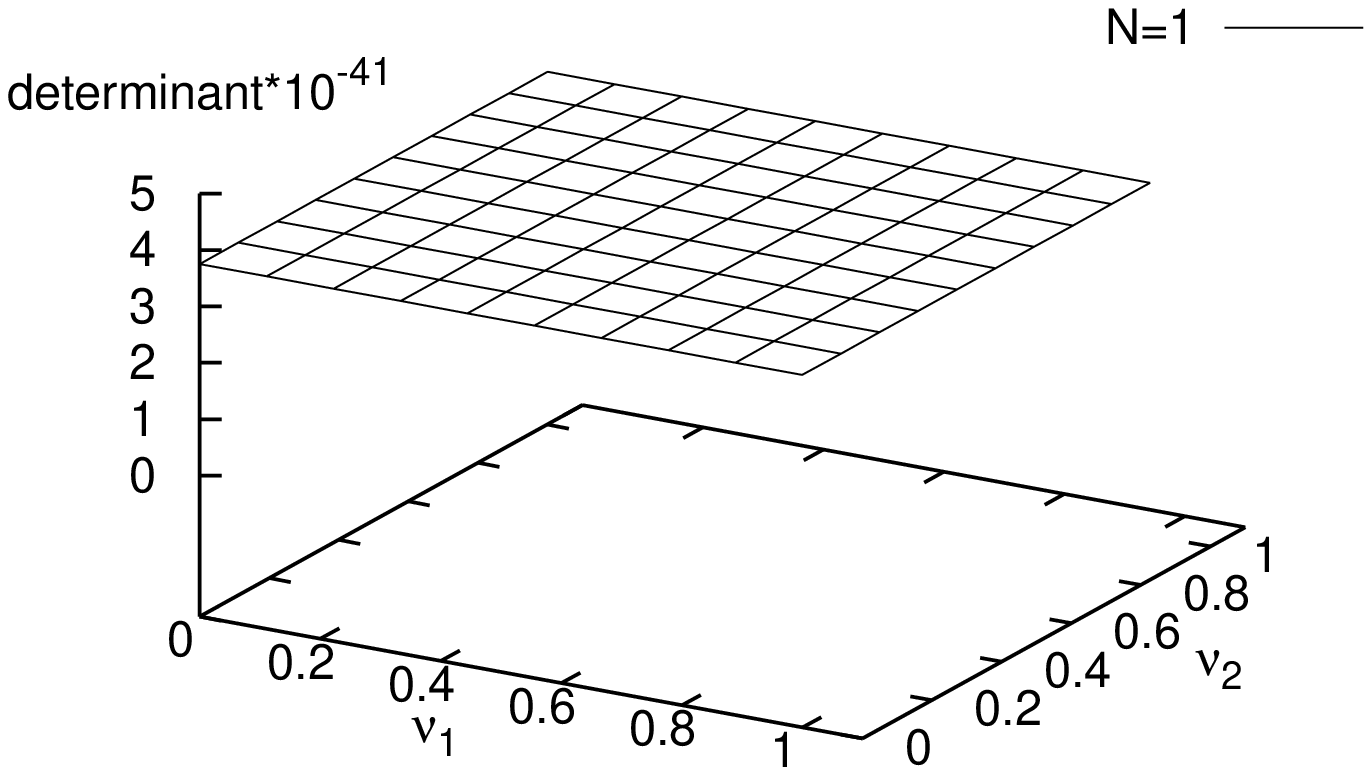}
\caption{Three dimensional plot of the $\nu$ dependence 
of the fermion determinant $\det(D^{N}_{DW})^{2}/\det(D^{N}_{AP})^{2}$. 
Left: $N=0$ case, Right: $N=1$ case.}
\label{fig:detm02n01}
\end{figure}

\begin{figure}[t]
\begin{minipage}{.45\linewidth}
\includegraphics[height=8cm,angle=-90]{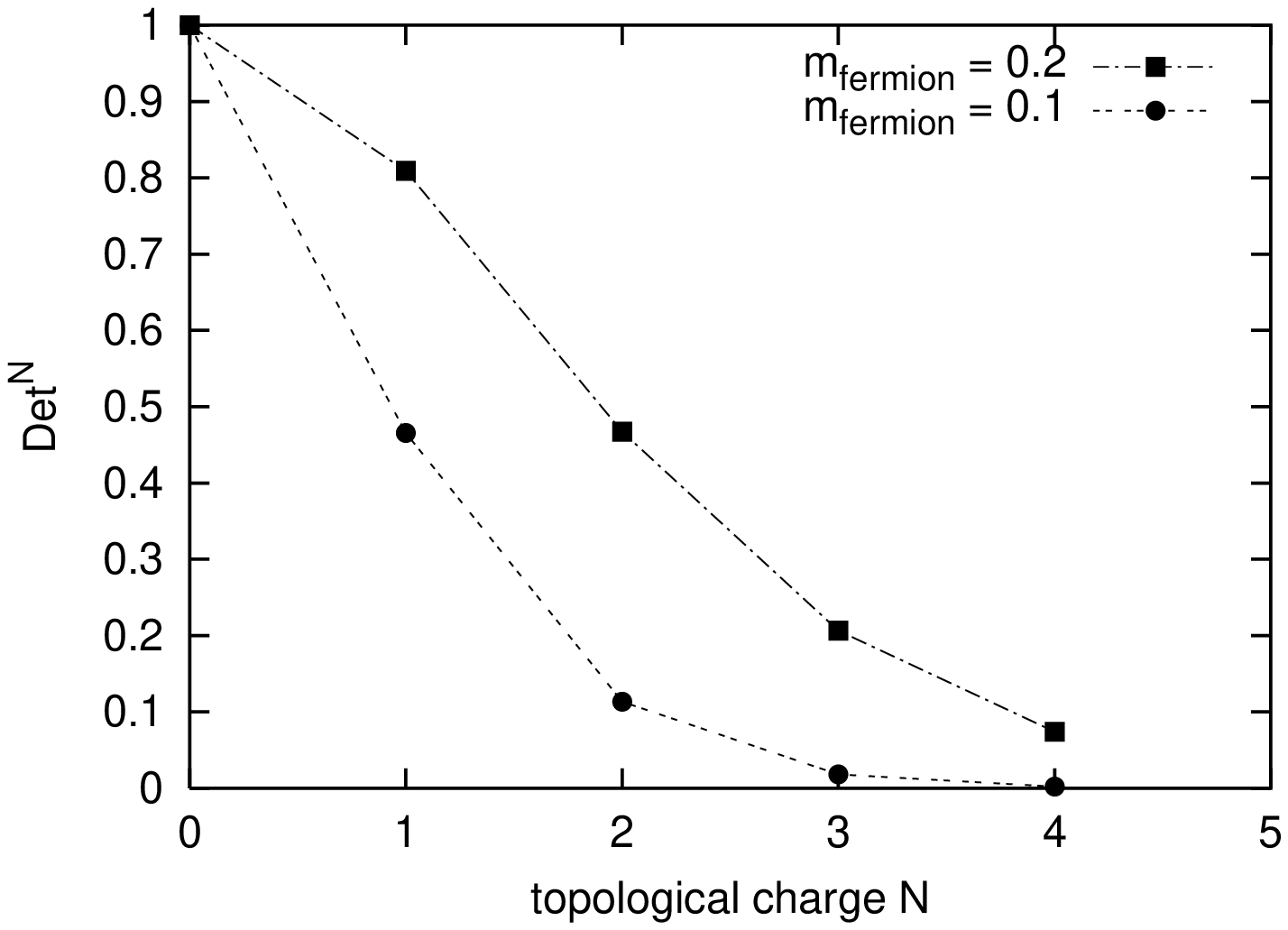}
\caption{$|N|$dependence of $Det^{N}$ for the fermion mass 
m = 0.1, 0.2 are shown.}
\label{fig:free det}
\end{minipage}
\begin{minipage}{.45\linewidth}
\includegraphics[height=8cm,angle=-90]{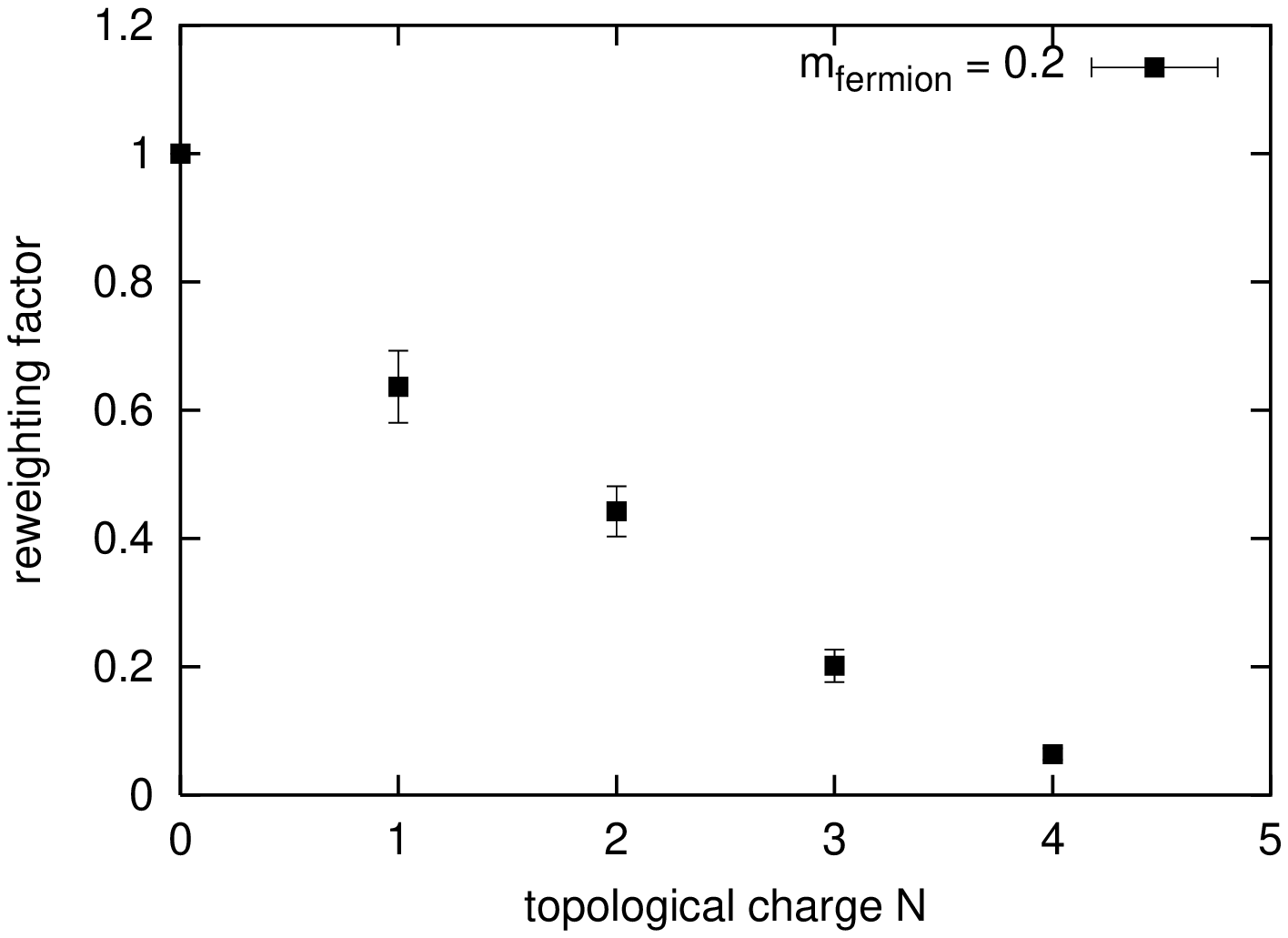}
\caption{The total reweighting factor $R^{N}(0.5,0.2)$ 
is plotted as a function of the topological charge. The
factor falls off rapidly as the topological charge increases.}
\label{fig:reweight}
\end{minipage}
\end{figure}

The fermion determinant $\det D^{2}$ on classical background in the 
sector with $N$ is numerically 
calculated using the Householder method and the QL method \cite{Numerical Recipe}.
The integral over the moduli $\nu_{1,2}$ is approximated by the 
weighted sum over the discrete set of points uniformly distributed 
in the whole integration region as Fig.~\ref{fig:detm02n01}. 
The numbers of points for the weighted sums 
are $5 \times 5$  for both $\det(D^{0}_{DW})^{2}/\det(D^{0}_{AP})^{2}$, 
and for $\det(D^{N}_{DW})^{2}/\det(D^{N}_{AP})^{2}$ with $N\neq 0$. 
The value of 
\begin{equation}
Det^{N} \equiv 
\frac{\int d\nu_1 d\nu_2 \det(D^{N}_{DW})^{2}/\det(D^{N}_{AP})^{2}}
     {\int d\nu_1 d\nu_2 \det(D^{0}_{DW})^{2}/\det(D^{0}_{AP})^{2}},
\end{equation}
is plotted in Fig.\ref{fig:free det}.
It decreases as $|N|$ increases,
due to the contribution of small eigenvalues proportional to the fermion
mass, which emerge in the nontrivial topological sectors 
since Atiyah-Singer index theorem is realized on the lattice 
in $L_{3}\to \infty$ limit \cite{Luscher:1998pq}.

In order to obtain the exponential factor in  
Eq.(\ref{eq:reweight}),  we need to evaluate the integral 
of the following quantity
\begin{equation}S_{subtr}^{N}(\beta^{\prime},m)\equiv
\langle  S_{G} - S^{N}_{G\;min}\rangle^{N}_{\beta^{\prime},m}
-\langle  S_{G} \rangle^{0}_{\beta^{\prime},m}.
\end{equation}
Since $S_{subtr}^{N}(\beta^{\prime},m)$ 
decreases rapidly as $\beta^{\prime}\to \infty$, 
the integral of $S_{subtr}^{N}$ over $\beta^{\prime}$ 
is well approximated by a weighted sum over the discrete set of points 
for $S_{subtr}^{N}(\beta^{\prime},m)$ at 
$\beta^{\prime}=0.5,1.0,1.5,2.0$.
For each $\beta^{\prime}$, we evaluate $S_{subtr}^{N}(\beta^{\prime},m)$
by sampling more than 5000 configurations.

Total reweighting factor $R^{N}(\beta,m)$ at $\beta=0.5$ and
$m=0.2$ is plotted in Fig.~\ref{fig:reweight}. It is shown that 
higher topological sectors are indeed suppressed by 
the reweighting factor.

Finally, combining the correlators and the reweighting factors, 
we obtain the  total expectation values on nonzero-$\theta$ vacuum
as 
\begin{equation}
\sum_{y}\langle \pi(x,y)\pi(0,0)\rangle_{full} =
\sum_{N=-4}^{4}e^{iN\theta}
\sum_{y}\langle \pi(x,y)\pi(0,0)\rangle^{N}_{\beta,m}
R^{N}(\beta,m),
\end{equation}
and
\begin{equation}
\sum_{y}\langle \eta(x,y)\eta(0,0)\rangle_{full} = 
\sum_{N=-4}^{4}e^{iN\theta}
\sum_{y}\langle \eta(x,y)\eta(0,0)\rangle^{N}_{\beta,m}
R^{N}(\beta,m),
\end{equation}
up to a constant normalization factor. 
Here we have ignored $|N| > 4$ sectors since they only give 
contributions less than 1.2 \% of zero sector.
Then we can get
pion mass and $\eta$ meson  mass including full nonperturbative effects and
$\theta$ dependence.
In this calculation, the propagators are fitted by minimizing the
$\chi^2$ and the total statistical errors are estimated by summing 
those in individual sections in quadrature.


\newpage
\subsection{Systematic errors}

In this subsection we discuss possible systematic errors.
These error estimations show that our simulation is 
reasonable and the results are reliable.  

\begin{figure}[b]
\includegraphics[height=9cm,angle=-90]{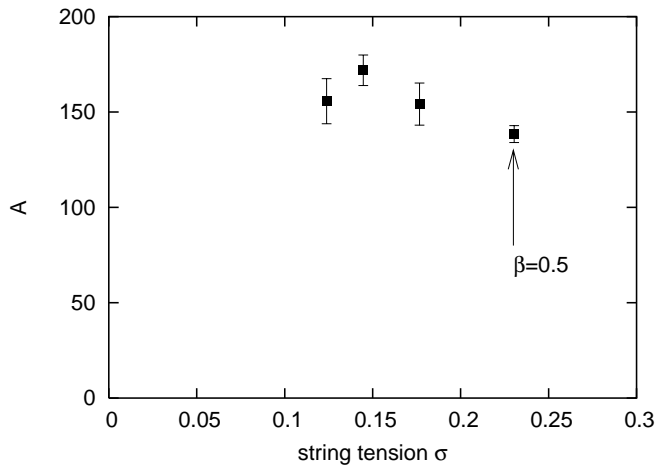}
\caption{Lattice spacing dependence of the dimensionless 
quantity $A\equiv m_{\pi}^{6}/(m^{4}\sigma)$. Horizontal 
axis is the string tension in lattice unit.} 
\label{fig:A_tension}
\end{figure}

Let us now study the lattice spacing dependence.
We measure a dimensionless quantity $A\equiv m_{\pi}^{6}/(m^{4}\sigma)$
at $\beta = 0.5,1.0,1.5,2.0$ in zero topological sector,
where $m_{\pi}$ denotes pion mass and $\sigma$
denotes the string tension. 
As Fig.~\ref{fig:A_tension} indicates, 
the results at $\beta=0.5$ show no large lattice spacing dependence
which suggests that the discretization error is under control.

\begin{figure}[t]
\begin{minipage}{.45\linewidth}
\includegraphics[height=8cm,angle=-90]{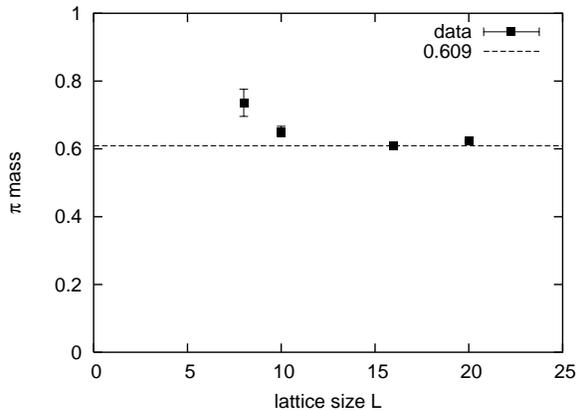}
\caption{The dependence of the pion mass on the lattice size $L$ 
in the space-time direction. The fermion mass is $m=0.2$.
Filled symbols are the data and the dotted line shows the fitted mass 
for $L=16$. The pion mass shows no volume dependence for $L \geq 16$. }
\label{fig:pi_size}
\end{minipage}
\begin{minipage}{.45\linewidth}
\includegraphics[height=8cm,angle=-90]{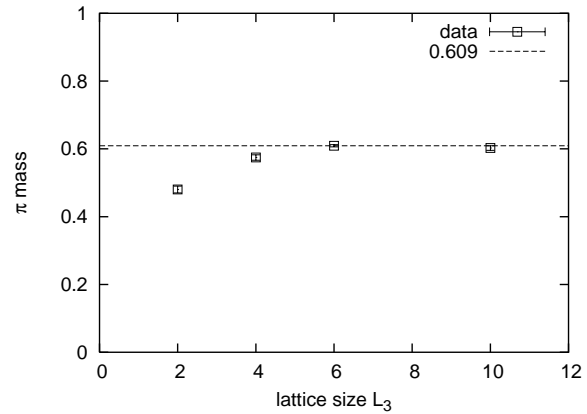}
\caption{The dependence of the pion mass on the lattice size $L_{3}$ 
in the third direction. The fermion mass is $m=0.2$.
Open symbols are the data and the dotted line shows the fitted mass 
for $L_3=6$. The pion mass shows no volume dependence for $L_3 \geq 6$. }
\label{fig:pi_dwsize}
\end{minipage}
\end{figure}

Next we discuss finite size effects for the space-time size $L$ and 
and for the extra dimension size $L_{3}$.
We measure the pion mass  on the lattices of size 
$L^{2} \times L_{3} = 8^{2}\times 6,10^{2}\times6,
16^{2}\times 6 ,20^{2}\times 6 ,
16^{2}\times 2,16^{2}\times 4$ and $16^{2}\times 10$ 
in the zero sector. 
Fig.~\ref{fig:pi_size} shows $L$ dependence and
Fig.~\ref{fig:pi_dwsize} shows $L_{3}$ dependence.
We find the meson mass is stable for $L$ larger than 16 
and for $L_3$ larger than 6 so that the finite size error is 
also under control with our choice of the lattice size $16^{2}\times 6$.
The discretization error and finite size errors from 
the nonzero topological sector is similarly under control.


We now study the error in the integration over the moduli
$\nu_{1,2}$. 
In order to estimate the systematic error we also evaluate
the integral by the weighted sum of $10 \times 10$ points. We find 
that the change is very tiny ( relative change $\sim 10^{-8}$ ) 
and is negligible compared to other systematic errors, as 
is expected from the mild $\nu$ dependence of 
$\det(D^{0}_{DW})^{2}/\det(D^{0}_{AP})^{2}$ in
Fig.~\ref{fig:detm02n01}. 
Fig.~\ref{fig:detm02n01} also shows that 
$\det(D^{N}_{DW})^{2}/\det(D^{N}_{AP})^{2}$ with $N\neq 0$
has almost no $\nu$ dependence. In fact this remarkable 
flat dependence is also seen in the continuum counterpart
analytically \cite{Azakov:1996xk}. We therefore conclude 
that the error in the weighted sum is even more negligible for 
the nonzero topological sector.

\begin{figure}[t]
\includegraphics[height=9cm,angle=-90]{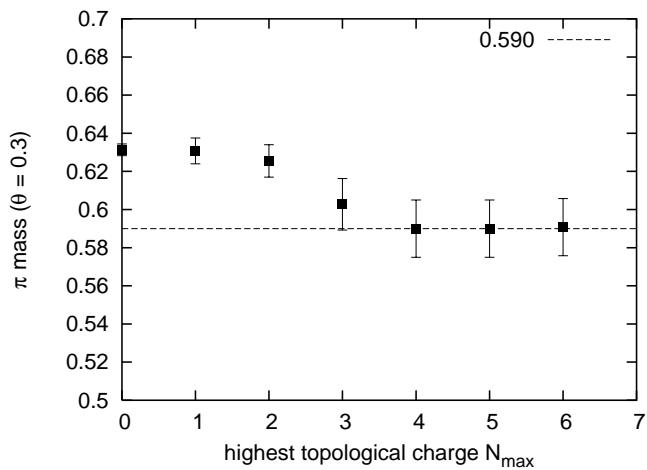}
\caption{The pion mass from the sum over $|N| \leq |N_{max}|$ sector 
contributions 
for $m=0.2$ and $\theta=0.3\pi$. No change in the mass 
for $N_{\rm max} \geq 4$ is observed. }
\label{fig:topmax}
\end{figure}

Since the integral of $S_{subtr}^{N}$ over $\beta^{\prime}$  is approximated 
by the trapezoidal rule using the data for the discrete
set of $\beta^{\prime}$ points, the error in this approximation should be
estimated. For this purpose we evaluate the integral of
$S_{subtr}^{N}$ 
in an alternative way, in which we fit the discrete set of data 
with the function of the form 
\begin{equation}
S_{subtr}^{N}(\beta^{\prime},m) = 
\frac{a_{1}}{\beta^{\prime \;2}}+\frac{a_{2}}{\beta^{\prime \;3}},
\end{equation}
and compute the integral of $S_{subtr}^{N}$ over $\beta^{\prime}$ analytically.
Table \ref{tab:R} shows 
the difference of the two ways of evaluation.
The pion mass are consistent with each other.
Thus we find that the approximation for the integral of
$S_{subtr}^{N}$ 
does not give large systematic errors in the meson mass.
\begin{table}
\begin{tabular}{ccc}
\hline
\hline
& By trapezoidal rule & By fit \\ \hline
$R^{0}(0.5,0.2)$ & 1.0 & 1.0 \\
$R^{1}(0.5,0.2)$ & 0.637(56)& 0.59(22)\\
$R^{2}(0.5,0.2)$ & 0.442(39)& 0.45(16)\\
$R^{3}(0.5,0.2)$ & 0.201(25)& 0.32(18)\\
$R^{4}(0.5,0.2)$ & 0.0636(91)& 0.072(46)\\
\\
$m_{\pi}$(at $\theta=0$) & 0.647(07) & 0.650(34) \\ \hline
\end{tabular}
\caption{The reweighting factor in each sector from two different
methods of evaluating the integral of $S_{subtr}^{N}$; 
The trapezoidal rule and the integral of the polynomial fit.  
The resulting pion masses are also given.  }
\label{tab:R}
\end{table}
We now study the truncation error in the sum over topological
sectors. As we discussed before, we neglect $|N|>4$ sectors
since these contribution are suppressed by 
large value of action and fermion zero modes.
Fig.~\ref{fig:topmax} shows the pion mass at $\theta=0.3\pi$ measured 
for a variety of the highest topological charge $N_{max}$.
Therefore the truncation error in the sum over topological 
sectors are negligible in comparison with the statistical errors.


\newpage

\section{Meson masses}\label{sec:Results}

\subsection{Pion mass and $\theta$ dependence}

\begin{figure}[b]
\includegraphics[height=10cm,angle=-90]{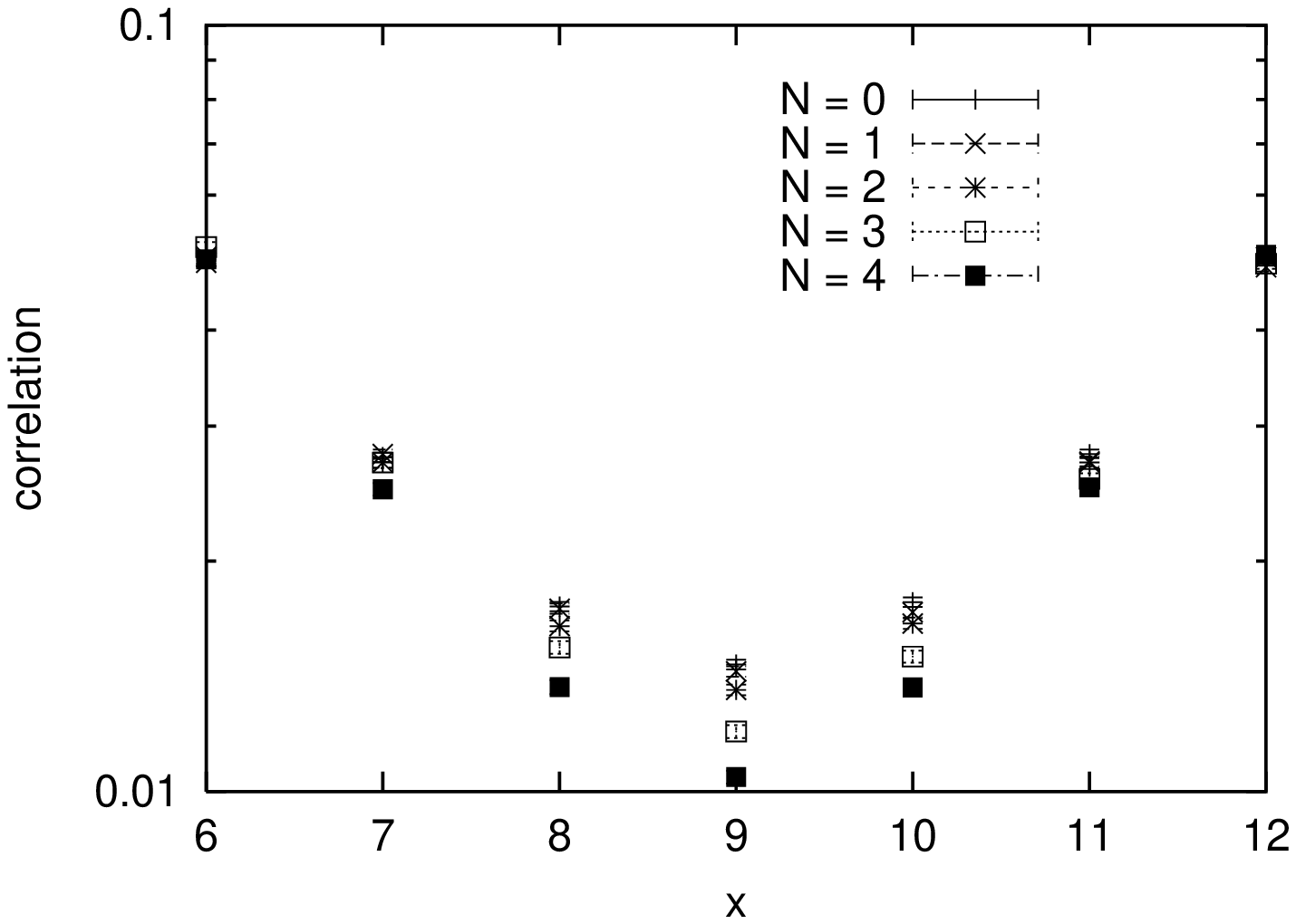}
\caption{The pion propagator in each sector for $m=0.2$.}
\label{fig:piprop1}

\begin{minipage}{.45\linewidth}
\includegraphics[height=8cm,angle=-90]{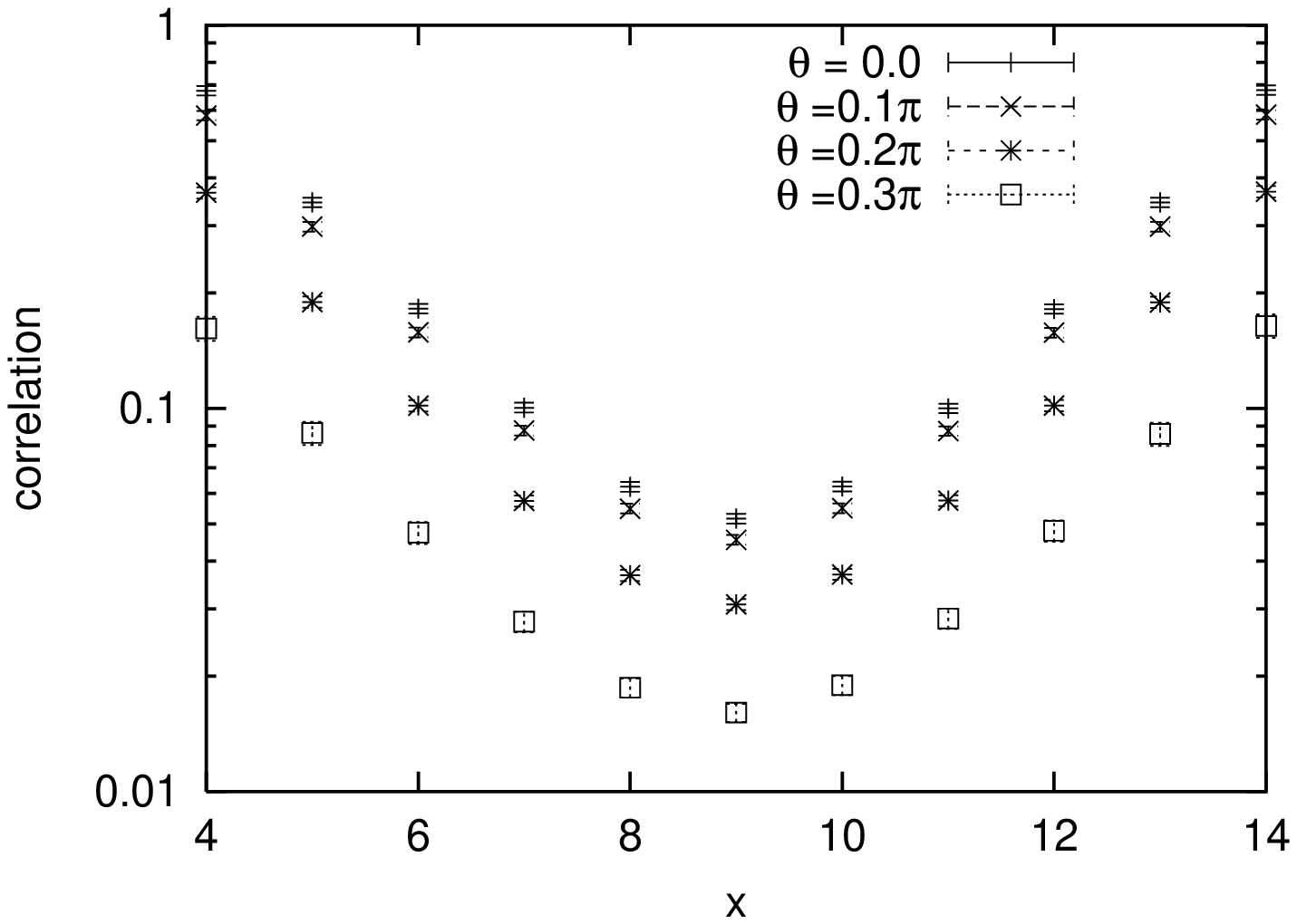}
\caption{The full pion propagators with $m=0.2$ for various $\theta$ 
are plotted.}
\label{fig:piprop2}
\end{minipage}
\begin{minipage}{.45\linewidth}
\includegraphics[height=8cm,angle=-90]{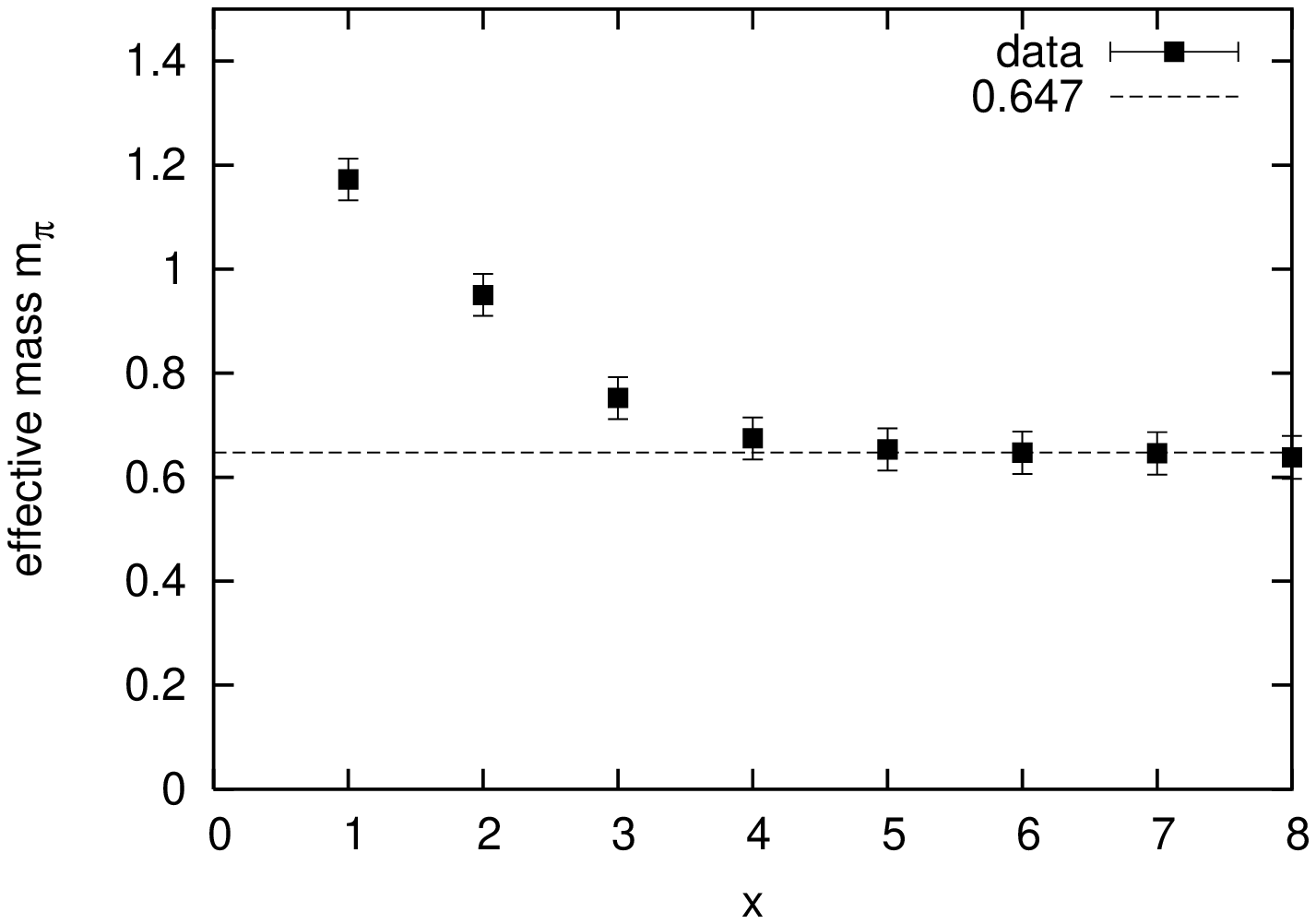}
\caption{The effective mass plot of the pion for $m=0.2$
and $\theta=0$. The dotted line shows the result of the fit.}
\label{fig:pieffmass}
\end{minipage}
\end{figure}

Fig.~\ref{fig:piprop1} shows pion propagators in each topological sector and 
Fig.~\ref{fig:piprop2} shows full propagators at various $\theta$.
We measure the pion mass by fitting these data to the 
hyperbolic cosine function. The fit range is $x =[5,8]$
for which we find good plateau in the effective mass plot 
as Fig.~\ref{fig:pieffmass} shows.
In fitting $\chi^{2}/dof$ is also a small value ($\chi^{2}/dof < 0.1$).

\begin{figure}[p]
\includegraphics[height=12cm,angle=-90]{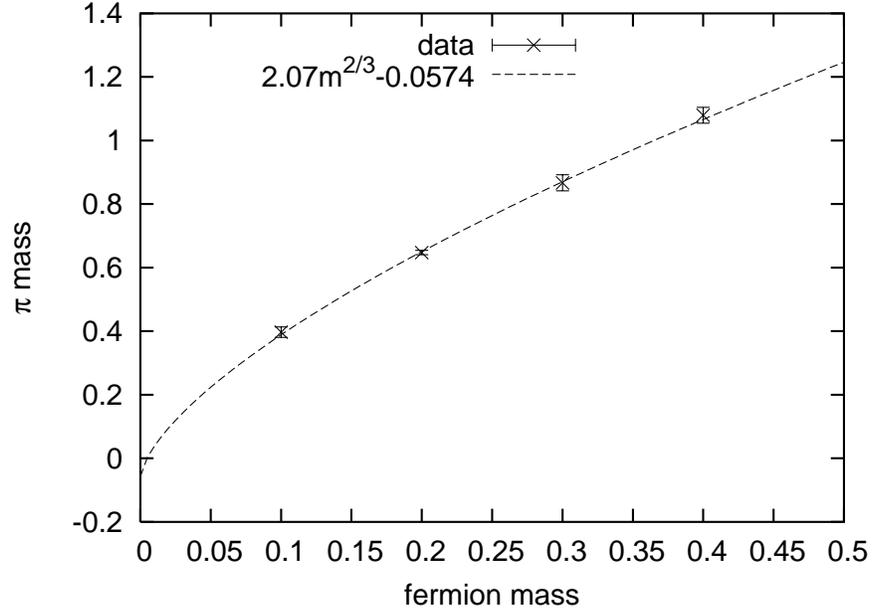}
\caption{The fermion mass dependence of the pion mass for $\theta=0$. 
The crosses are the lattice data and the dotted line is the result 
of the fit with the function in Eq.(\ref{eq:chiral}).
The chiral behavior is consistent with that of continuum theory.}
\label{fig:pimass}
\end{figure}
\begin{figure}[p]
\includegraphics[height=12cm,angle=-90]{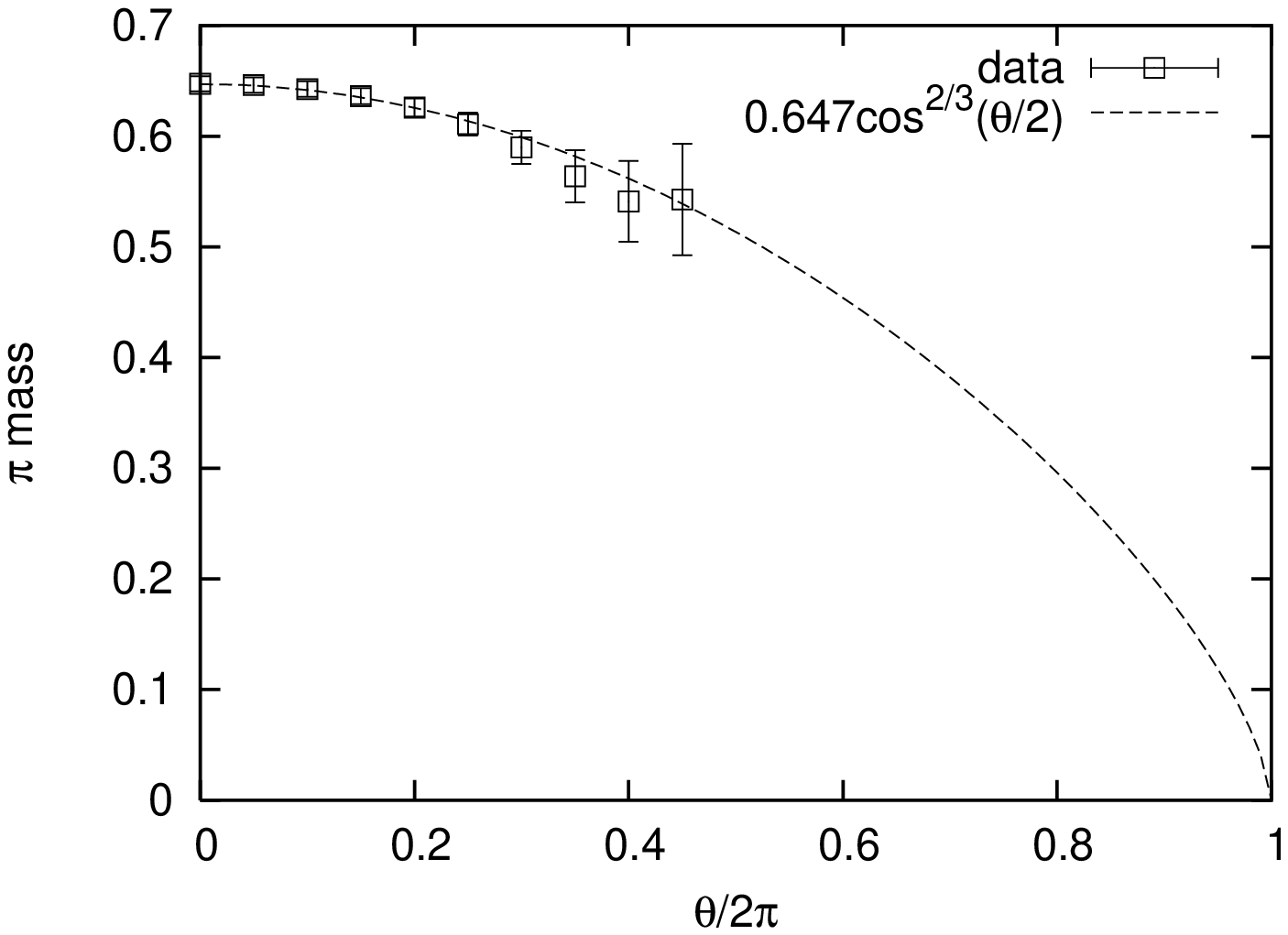}
\caption{$\theta$ dependence of the pion mass at $m=0.2$.
The open symbols are the lattice data. 
The dotted line is the analytical result of the $\theta$ dependence 
in the continuum theory, where the normalization is fitted by the 
lattice results. For $\theta/(2\pi) < 0.5 $, the pion mass 
is proportional to $\cos (\theta/2)^{2/3}$, which is in complete
agreement with the continuum results.}
\label{fig:pitheta}
\end{figure}

Fig.\ref{fig:pimass} shows pion mass at $\theta =0$
as a function of fermion mass $m$.
We ignore the $m$ dependence of $S_{subtr}^{N}(\beta^{\prime},m)$ 
and use $m=0.2$ result for all $m$. 
We fit the results to the following function suggested by 
the continuum theory with possible additional constant term $b$ from
the residual mass of pion,
\begin{eqnarray}
m_{\pi}(m) &=&  am^{2/3}+b. 
\label{eq:chiral}
\end{eqnarray}
Fig.\ref{fig:pimass} shows that Eq.~(\ref{eq:chiral}) fits the data 
very well ($\chi^{2}/dof=0.39$) so that the fermion mass dependence is 
consistent with the continuum theory. The residual mass of the pion 
measured in the chiral limit is also tiny as $b =  -0.057 \pm 0.060 $,
which shows that the violation of the chiral symmetry is very small.

In Fig.~\ref{fig:pitheta} we present the $\theta$ dependence of the 
pion mass at $\beta=0.5$ and $m=0.2$.
As a remarkable feature, the result is in perfect agreement with that 
in the continuum theory in $\theta/(2\pi) < 0.5$ region.
A good control of the $\theta$ dependence shows 
that our method for summing over different topological sectors 
 with L\"uscher's gauge action indeed works numerically.

\begin{figure}[t]
\includegraphics[height=9cm,angle=-90]{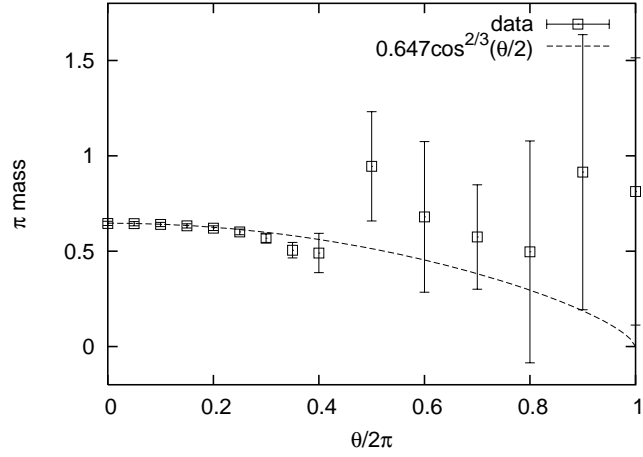}
\caption{$\theta$ dependence of the pion mass obtained 
by ignoring $S_{subtr}^{N}(\beta^{\prime},m)$ in the reweighting 
factors. }
\label{fig:pithetaquench}
\end{figure}

At large $\theta$, statistical errors increase, which 
are due to cancellations of propagators among different topological 
sectors.
In the calculation, we approximate the integral 
of $S_{subtr}^{N}(\beta^{\prime},m)$ by the trapezoidal rule 
for the discrete set of $\beta^{\prime}$ points, but this does not seem to 
be the reason for the large fluctuation in the $\theta/(2\pi) > 0.5$ region.
The main nonperturbative 
contribution comes from $Det^{N}$ and
$S_{subtr}^{N}(\beta^{\prime},m)$ gives only 
perturbative effects of order $\beta^{\prime -2}$.
In fact, even ignoring the integral (we set 
$S_{subtr}^{N}(\beta^{\prime},m)=0$ for all $\beta^{\prime}$)
we can get similar results as Fig.~\ref{fig:pithetaquench}.

We suspect that this large fluctuation is an example of the 
well-known phase problem.
Simply increasing the statistics might not improve the
situation.

Of course in application to QCD, it will be important
to evaluate $S_{subtr}^{N}(\beta^{\prime},m)$ 
and other observables more precisely. 


\newpage
\subsection{$\eta$ meson correlator and U(1) problem}

As the final subject, we would like to present the result of 
our exploratory measurement of the $\eta$ meson mass in order to
study the topological structure.
The $\eta$ propagator consists of two parts;
\begin{equation}
\langle\eta\eta\rangle = 
-2\langle tr\left(\gamma_{3}\frac{1}{D}\gamma_{3}\frac{1}{D}\right)\rangle
+4\langle tr\left(\gamma_{3}\frac{1}{D}\right)
tr\left(\gamma_{3}\frac{1}{D}\right)\rangle,
\end{equation}
where the first term is the same as flavor non-singlet 
$\pi$ propagator and the second term gives the ``hair-pin'' 
or disconnected contribution to the flavor singlet operator.
Because the number of physical space-time points is only
$16\times 16$, we compute the ``hair-pin'' contribution by brute
force, namely by solving the fermion propagator for all points
without relying on the noise method\cite{ref:noise}
or Kuramashi method \cite{ref:kuramashi}.

\begin{figure}[b]
\begin{minipage}{.45\linewidth}
\includegraphics[height=8cm,angle=-90]{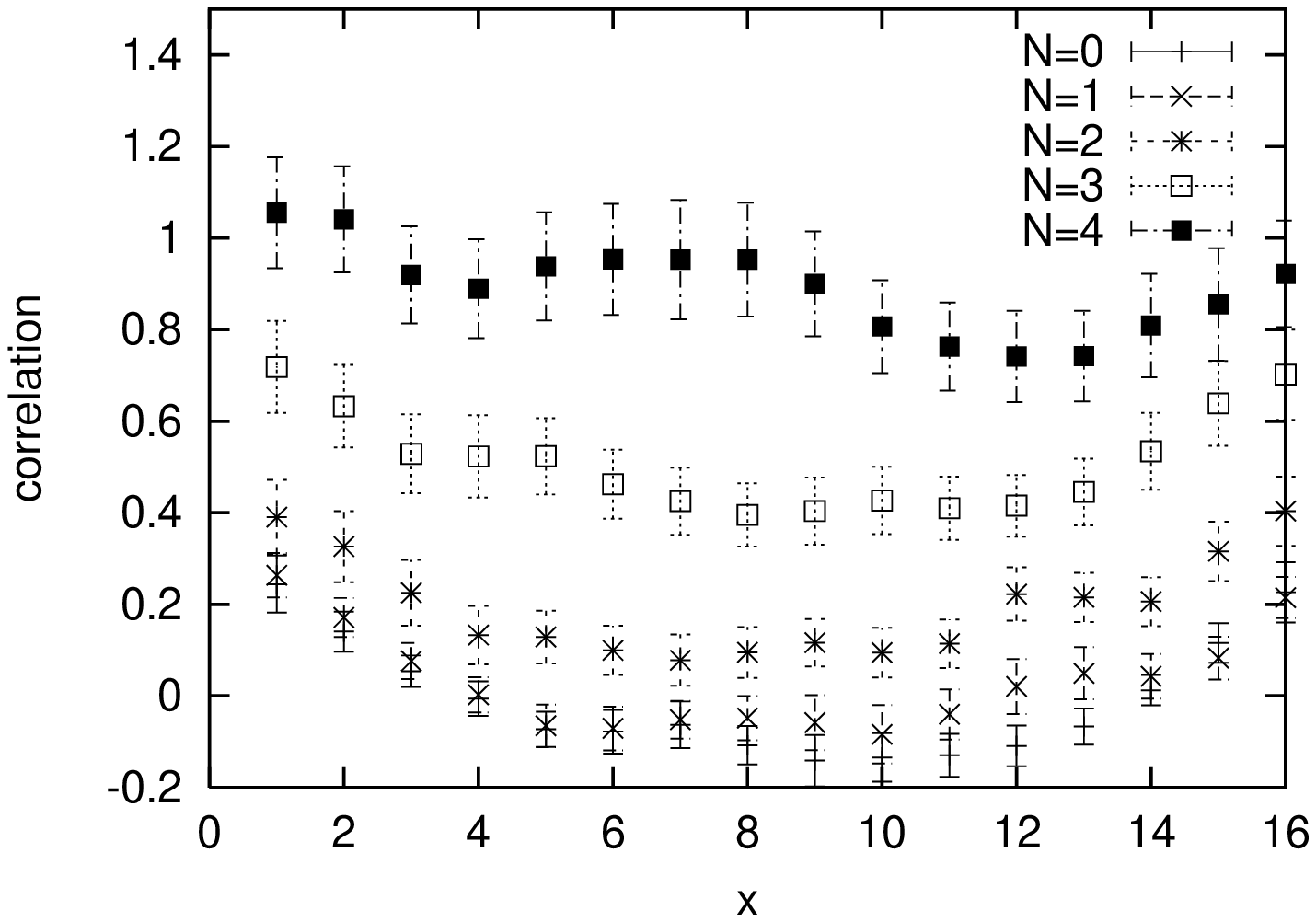}
\caption{The propagator of $\eta$ in each sector at $m=0.2$.}
\label{fig:etapropsec}
\end{minipage}
\begin{minipage}{.45\linewidth}
\includegraphics[height=8cm,angle=-90]{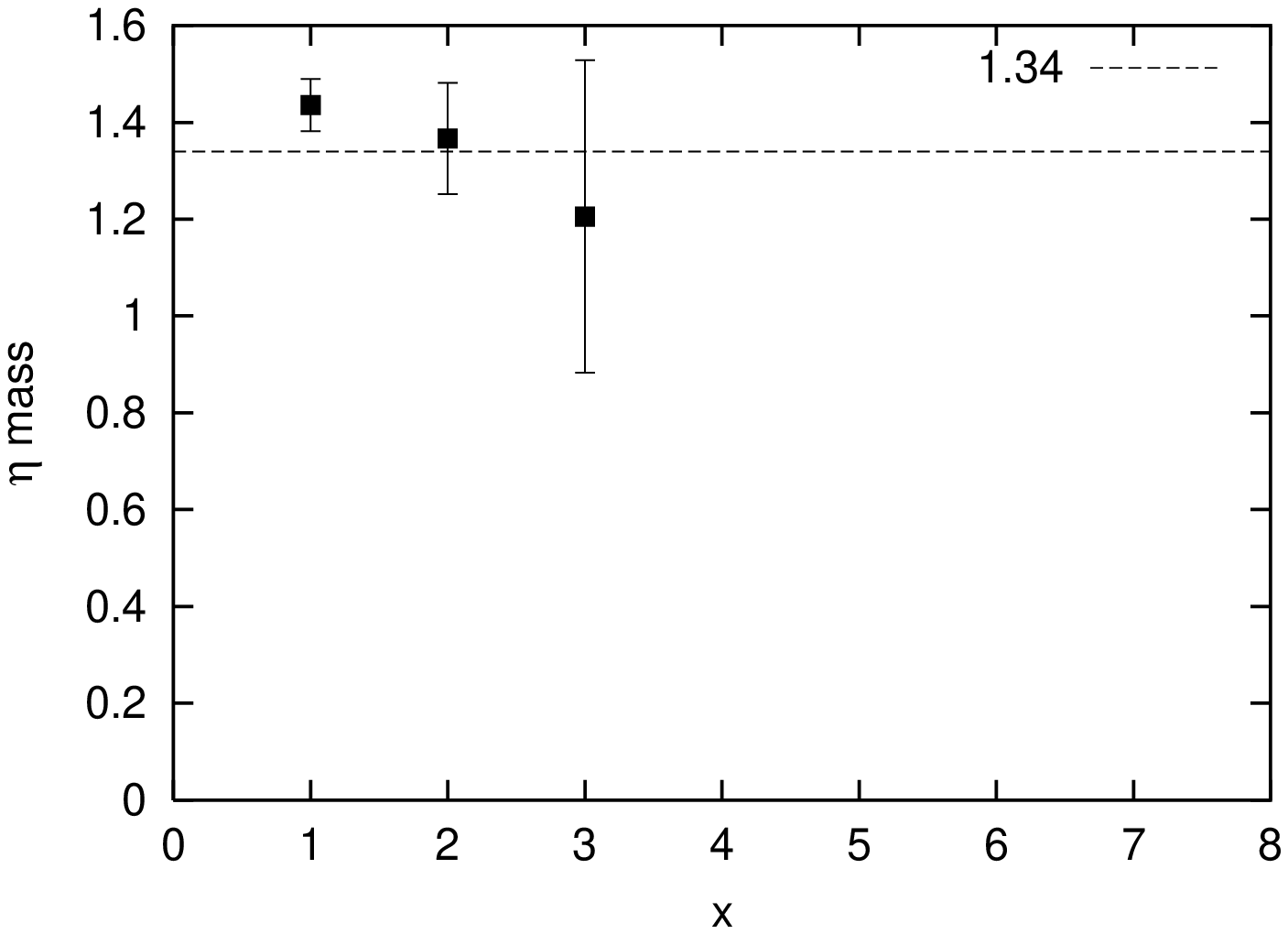}
\caption{The effective mass plot of the full $\eta$ propagator for
$m=0.2$, $\theta=0$.}
\label{fig:etaeffmass}
\end{minipage}
\end{figure}

\begin{figure}[t]
\includegraphics[height=10cm,angle=-90]{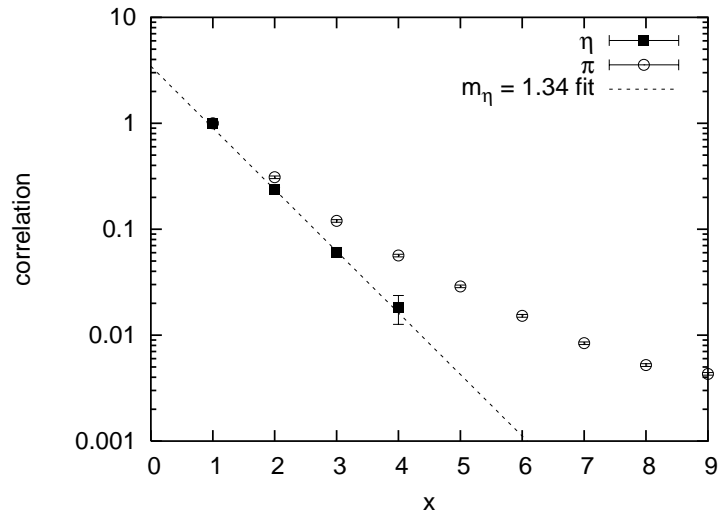}
\caption{The full propagator $\eta$ at $m=0.2$ and $\theta=0$ 
( filled squares ). The pion propagator is also plotted 
for comparison ( open circles ). 
The propagators are normalized by the value at $x=1$.}
\label{fig:etapi}
\end{figure}

Fig.~\ref{fig:etapropsec} shows the contribution of the second term 
in each sector, whereas  Fig.~\ref{fig:etapi}
shows the full (symmetrized) $\eta$ propagator at $m=0.2$ and $\theta=0$.
We also present effective mass plot in 
Fig.\ref{fig:etaeffmass}.
We find that the fall of $\eta$ propagator is steeper than that 
of $\pi$ which gives qualitatively consistent results with $U(1)$ problem,
although it suffers from large statistical fluctuations making 
quantitative studies difficult.
Further quantitative studies require some new ideas which efficiently
reduce the statistical error and controls the systematic errors due to
the summation over large number of topological sectors.

\section{Summary and Discussion}\label{sec:conclusion}

In this paper, we elucidate the role of the admissibility condition 
on the topological and chiral properties in lattice gauge theories
by applying L\"uscher's action  together with domain wall fermions to 
a numerical simulation of the massive Schwinger model. 
To investigate the $\theta$-dependence of the correlators,
we have developed a method to sum over different topological 
sectors.  We have found that L\"uscher's action is indeed applicable 
to Monte Carlo simulations and all the results are consistent with 
those in the continuum theory, confirming the validity of our method.

We summarize the features of this action here again.
(1) In L\"uscher's action, the gauge field strength is uniquely 
determined from the plaquette and the gauge action is a smooth 
function of the field strength.
(2) The range of the action is not compact;
       \begin{equation}
	0 \leq S_{G} < \infty.
       \end{equation}
       This is the same situation as continuum theory.
We can treat the theory in terms of the field strength rather than 
plaquettes.
According to these features, L\"uscher's gauge  action  has many advantages.
\begin{enumerate}
\item  
The use of this gauge action with the domain 
wall fermion action is valid even for the strong coupling regime since
unphysical configurations are suppressed.
(We find the suppression effect is especially remarkable in quenched 
approximation as discussed in Appendix.)
\item 
We can treat the topological properties of 
the lattice theories precisely. 
This exact topological treatment is useful not only mathematically 
but also in a practical point of view. In conventional 
approach, there are two technical problems. i.e.
violation of chirality at strong coupling and the slowing down 
of the topology change in unquenched simulation. For the 
former problem, the improved gauge actions which 
suppress the dislocations is proposed. However, in principle
the suppression of the dislocations also suppresses the topology
change so that the latter problem becomes even more difficult.
Our method makes the improvement to the extreme and prohibits both 
the dislocation and the topology change completely, however by 
computing each topological sector and its reweighting factor 
we can reconcile the solutions to the the topology change problem 
and the dislocation problem at the same time.
\item 
Once each topological sector can be 
computed  separately, we can obtain a $\theta$ dependence at once. 
\item 
Aside from the fact that we must simulate for each sector 
the typical simulation time needed for the trivial 
topological sector is no larger than that of using Wilson's plaquette
action. For the nonzero topological charge sector, one can also 
increase the statistics at will very efficiently, in contrast to 
the conventional method where one can increase the statistics 
only by reaching the thermal equilibrium. 
In this sense, our method would have advantages in physical quantities 
for which the topological sectors with larger instanton numbers 
give larger contributions.
\end{enumerate}

We expect that it would not be difficult to apply L\"uscher's type of
gauge action also to QCD in four dimensions. In order to study 
the $\theta$ vacuum, one should study how the reweighting factors can be 
computed. Also one should study the exact topological index on the
lattice since the topological properties may be very complicated in 
4-dimensional QCD on a torus. 
At least we can recommend the application to the calculation
in the exact chiral limit since we usually need only zero sector results. 
We hope the understanding of the topological properties
in lattice QCD will be improved by applying L\"uscher's 
admissibility condition.

Finally we would like to point out that the method proposed in this
paper to use L\"uscher's type gauge action and sum over topological 
sectors would also be essential for the numerical simulations 
of the chiral gauge theories on the lattice in the future.

\section*{Acknowledgments}\label{sec:acknowlegments}

We acknowledge Hideo Matsufuru and Ayumu Sugita for the 
discussions and help throughout the whole stages of this work 
and Hideo Matsufuru also for his careful reading of the manuscript
and his comments.  We would also like to thank Yoshio Kikukawa for 
his beautiful lectures as well as private discussions on recent 
developments chiral gauge theories on the lattice.  
We are also grateful to Kenichi Shizuya, Masanori
Okawa, Yusuke Taniguchi, Yoshinobu Kuramashi, Yasumichi Aoki 
for useful discussions.
The authors thank the Yukawa Institute for Theoretical Physics at
Kyoto University, where this work was initiated during the
YITP-W-02-15 on ``YITP School on Lattice Field Theory''.
The numerical simulations were done on Alpha workstation at 
Yukawa Institute for Theoretical Physics in Kyoto 
University,  NEC SX-5 at Research Center for Nuclear Physics in 
Osaka University, and Hitachi SR8000 model F1 supercomputer at KEK.

\section*{Appendix}\label{ap:quenched result}

In this appendix, we examine the validity of L\"uscher's action 
in the quenched approximation.
In the strong coupling region, we compare 
L\"uscher's action with Wilson's action
ignoring fermion loops.
We set lattice size to be $32 \times 32 \times 5$
and measure pion mass in the zero sector.
Gauge coupling $\beta$ is chosen to give the
same string tension $\sigma =0.18$ ; $\beta=1.0$ for L\"uscher's action
and $\beta=3.4$ for Wilson's action.

\begin{figure}[hbtp]
\includegraphics[height=8cm,angle=-90]{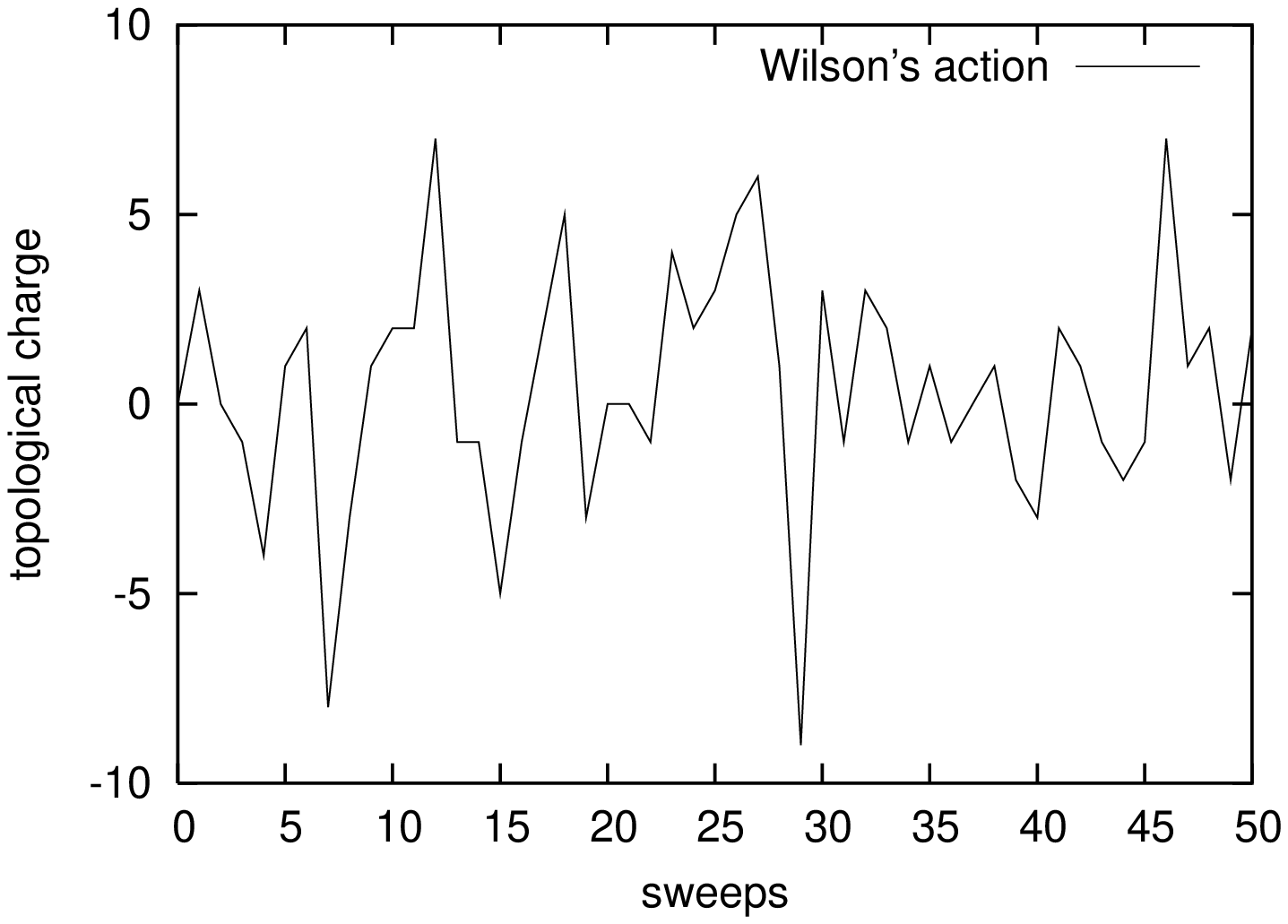}
\includegraphics[height=8cm,angle=-90]{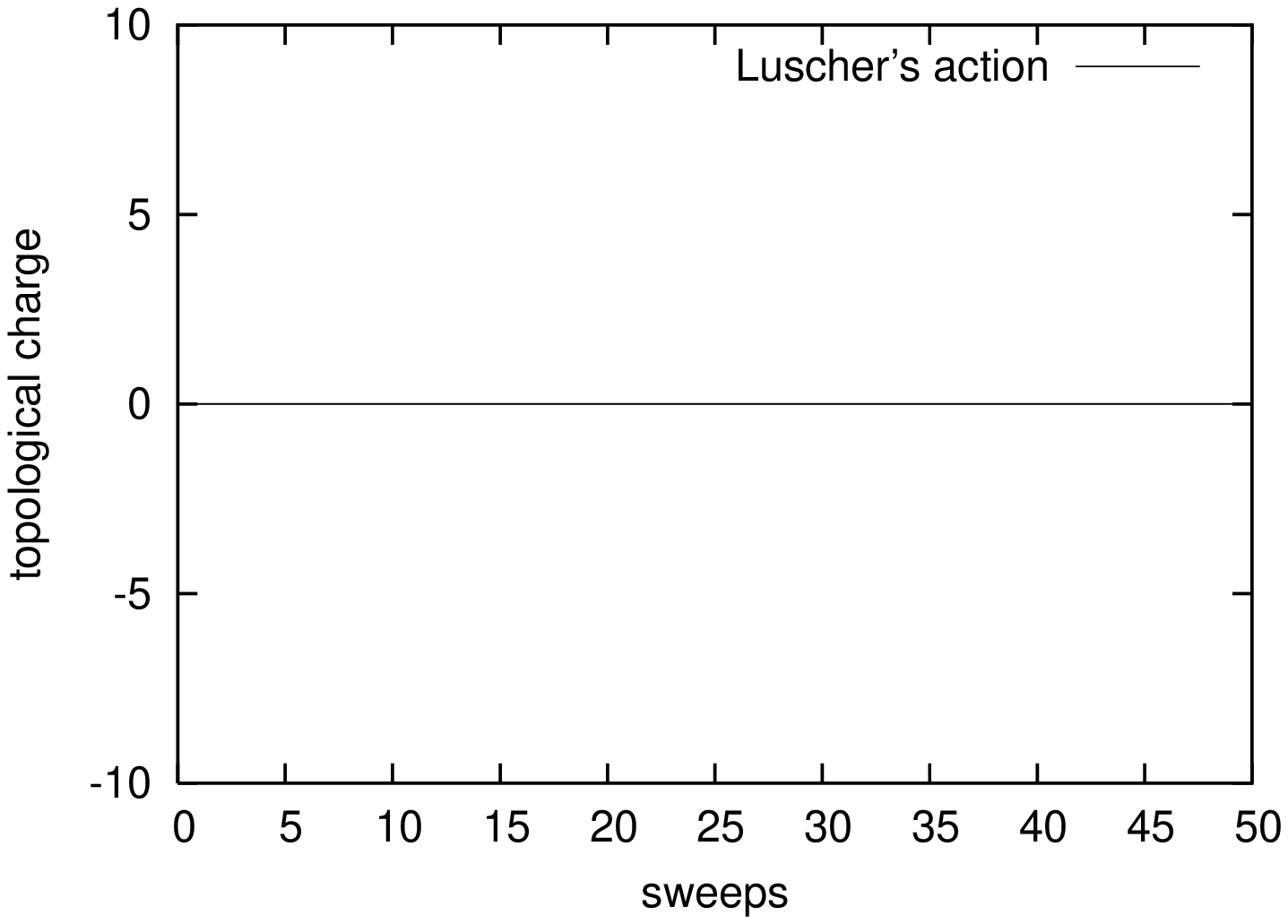}
\caption{The Monte Carlo evolutions of the topological charge 
in the quenched calculation with Wilson's gauge action and L\"uscher's 
gauge action for the gauge couplings having the same string tension. 
Left: Wilson's gauge action at $\beta=3.4$.  
Right: L\"uscher's gauge action at $\beta=1.0$. 
L\"uscher's gauge action shows no topology change.}
\label{fig:top changes}
\end{figure}

\begin{figure}[htbp]
\includegraphics[height=8cm,angle=-90]{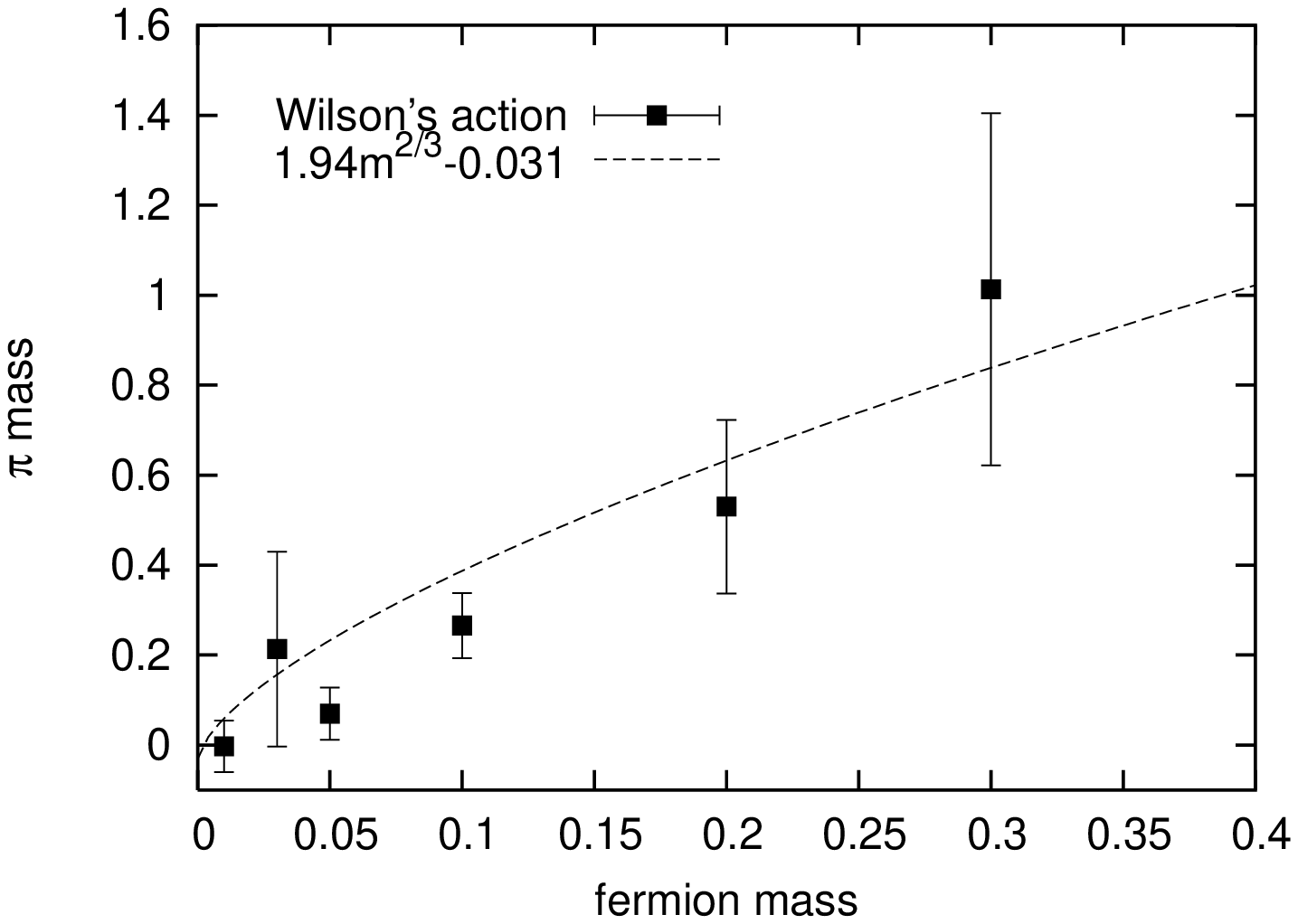}
\includegraphics[height=8cm,angle=-90]{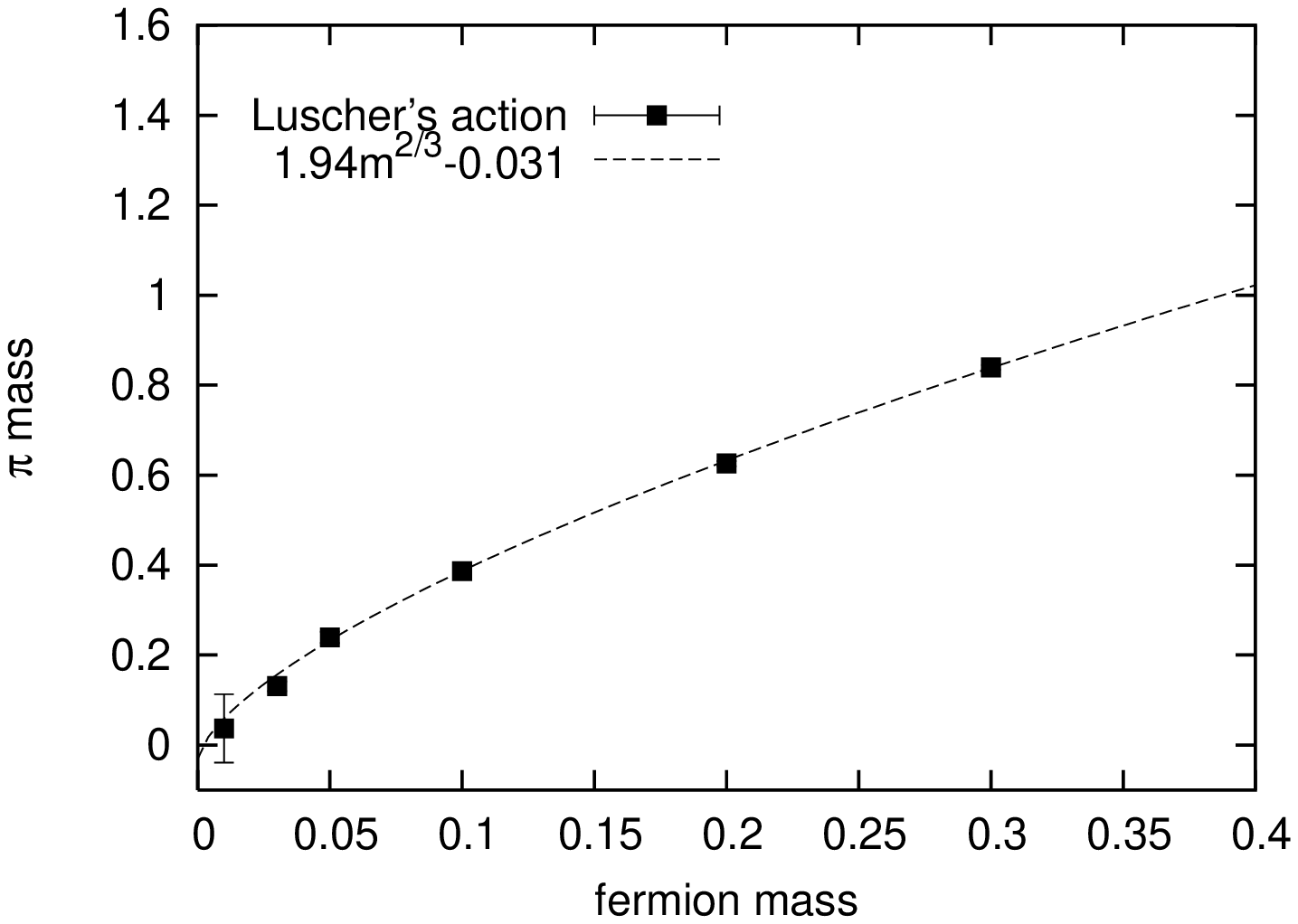}
\caption{The chiral behaviors of the pion mass in the quenched calculation 
with Wilson's gauge action and L\"uscher's gauge action for the gauge 
couplings having the same string tension. In this quenched study for 
illustrating the effect of the admissibility condition, only the 
sector with topological zero is taken for the calculation 
with L\"uscher's gauge action. 
Left: Wilson's gauge action at $\beta=3.4$ . 
 Right: L\"uscher's gauge action at $\beta=1.0$ . 
Wilson's gauge action suffers from large 
fluctuation while L\"uscher's gauge action shows a good chiral
behavior.
Both of them are calculated by domain-wall fermions.}
\label{fig:pimass quenched}
\end{figure}

Fig.\ref{fig:top changes} shows evolution of topological charge.
In quenched approximation with Wilson action, 
fermion zero modes are all neglected 
so there is no suppression on topology changes.
As a result, much unphysical configurations are 
generated.
On the other hand, L\"uscher's action never allow 
topology changes.
In Fig.\ref{fig:pimass quenched}, the difference is clear.
L\"uscher's action gives consistent pion mass and good
chiral limit even in quenched approximation at
very strong coupling.

For a theoretically complete study of the quenched Schwinger model, 
we should take a sum over different topological sectors.
However in our ``quenched'' study , unlike in the unquenched case, 
only the sector with zero topological charge is taken for the 
calculation with L\"uscher's gauge action, in order to illustrate 
the effect of the admissibility condition with the simplest example. 
It will be necessary to perform a theoretically complete quenched study 
by summing over different topologies and compare with analytic result
in which it is predicted that the quenching effect does give a 
different fermion mass dependence from that in the unquenched theory
\cite{ref:Sharpe}.


\newpage


\end{document}